\documentclass[twocolumn,english,reprint,amsmath,amssymb, onecolumn,nofootinbib,11pt]{revtex4-2}
\usepackage[T1]{fontenc}
\usepackage[latin9]{inputenc}
\makeatletter
\usepackage{float}
\usepackage{physics}
\usepackage{amsfonts}
\usepackage{latexsym}
\usepackage{epsfig}
\usepackage{graphicx}
\usepackage{tikz}
\definecolor{greatblue}{RGB}{40,120,181}
\definecolor{greatred}{RGB}{200,36,35}
\usepackage[colorlinks,linkcolor=greatblue,anchorcolor=blue,citecolor=greatred]{hyperref}

\usepackage{dcolumn}% Align table columns on decimal point
\usepackage{bm}% bold math

\makeatother

\usepackage[english]{babel}
\begin{document}
\preprint{preprintnumbers}{CTP-SCU/2025001}

\title{Mixed State Entanglement Entropy in CFT}

\author{Xin Jiang}  
\email{domoki@stu.scu.edu.cn}
\affiliation{College of Physics, Sichuan University, Chengdu, 610065, China}

\author{Peng Wang}
\email{pengw@scu.edu.cn}
\affiliation{College of Physics, Sichuan University, Chengdu, 610065, China}

\author{Houwen Wu}
\email{iverwu@scu.edu.cn}   
\affiliation{College of Physics, Sichuan University, Chengdu, 610065, China}

\author{Haitang Yang}
\email{hyanga@scu.edu.cn}
\affiliation{College of Physics, Sichuan University, Chengdu, 610065, China}

\begin{abstract}

How to calculate the entanglement entropy between two subsystems for a mixed state has remained an important problem. In this paper, we provide a straightforward method, namely the subtraction approach, to solve this problem for generic covariant bipartite mixed states, where time dependence is explicitly included.   We further demonstrate that the mixed state entanglement entropy  $S_\text{vN}$ can be calculated in a more abstract yet powerful way. Within the context of the AdS$_3$/CFT$_2$ and a configuration of AdS$_5$/CFT$_4$ correspondences, we show that  $S_\text{vN}$  exactly matches the corresponding entanglement wedge cross section in the AdS bulk, respectively. 
\end{abstract}
\maketitle
\newpage 
\tableofcontents

\section{Introduction}

Quantum entanglement has unveiled profound connections between quantum
gravity and quantum information theory. Within the framework of the
AdS/CFT correspondence
\citep{Maldacena:1997re,Witten:1998qj,Gubser:1998bc}, the entanglement
entropy \citep{Ryu:2006bv,Ryu:2006ef,Hubeny:2007xt} provides significant
insights into the relationship between entanglement and the emergence
of spacetime \citep{VanRaamsdonk:2010pw,Jiang:2024xcy,Jiang:2024xqz}.
The entanglement entropy, also known as the von Neumann entropy, is
defined as
\begin{equation}
S_{\mathrm{vN}}\left(A\right)=-\mathrm{Tr}\rho_{A}\log\rho_{A},
\end{equation}
where $\rho_{A}$ is the reduced density matrix of a subsystem $A$.
The entanglement entropy satisfactorily characterizes the quantum entanglement
between two subsystems in a \emph{pure state}. However, for the bipartite
systems in a \emph{mixed state}, such as two disjoint intervals in
CFT$_{2}$, the traditional entanglement entropy is no longer a suitable
measure of quantum entanglement.  Several approaches have been proposed to 
quantify the degree of entanglement in a mixed state,
such as the entanglement negativity \citep{Vidal:2002zz, Plenio:2005cwa,Calabrese:2012nk}, the
entanglement of purification \citep{Terhal:2002riz, Takayanagi:2017knl}, the odd
entanglement entropy \citep{Tamaoka:2018ned}, the reflected entropy
\citep{Dutta:2019gen}, and the balanced partial entanglement \citep{Wen:2021qgx}.
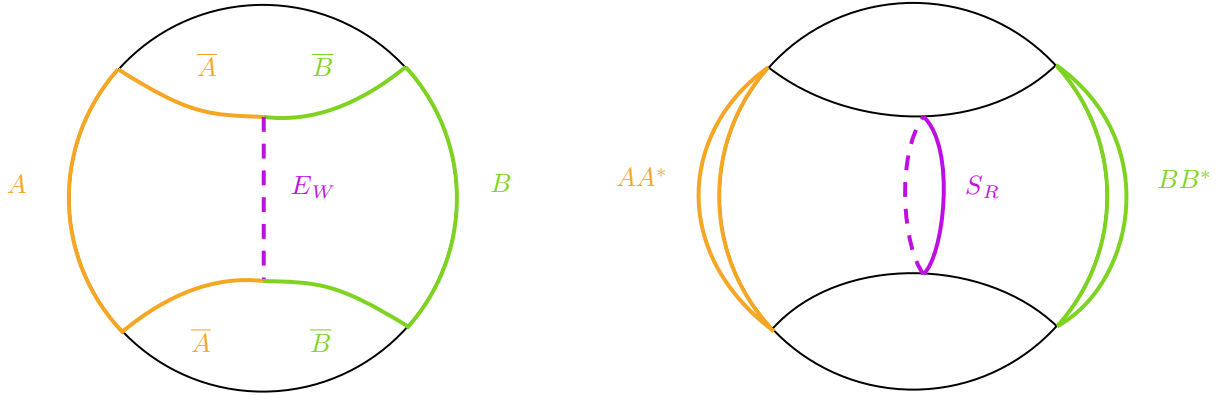
\begin{figure}[t]
\begin{centering}

\tikzset{every picture/.style={line width=0.75pt}} %set default line width to 0.75pt        

\begin{tikzpicture}[x=0.75pt,y=0.75pt,yscale=-1,xscale=1]
%uncomment if require: \path (0,515); %set diagram left start at 0, and has height of 515

%Shape: Ellipse [id:dp19969559243504564] 
\draw   (579.74,307.13) .. controls (579.74,253.23) and (623.43,209.53) .. (677.33,209.53) .. controls (731.23,209.53) and (774.93,253.23) .. (774.93,307.13) .. controls (774.93,361.03) and (731.23,404.72) .. (677.33,404.72) .. controls (623.43,404.72) and (579.74,361.03) .. (579.74,307.13) -- cycle ;
%Shape: Arc [id:dp4774767151953172] 
\draw  [draw opacity=0][line width=1.5]  (607.09,375.25) .. controls (590,357.63) and (579.48,333.61) .. (579.48,307.13) .. controls (579.48,282.13) and (588.86,259.32) .. (604.29,242.02) -- (677.33,307.13) -- cycle ; \draw  [color={rgb, 255:red, 245; green, 166; blue, 35 }  ,draw opacity=1 ][line width=1.5]  (607.09,375.25) .. controls (590,357.63) and (579.48,333.61) .. (579.48,307.13) .. controls (579.48,282.13) and (588.86,259.32) .. (604.29,242.02) ;  
%Shape: Arc [id:dp9582990093469455] 
\draw  [draw opacity=0][line width=1.5]  (749.35,240.88) .. controls (765.41,258.34) and (775.21,281.64) .. (775.18,307.23) .. controls (775.16,332.48) and (765.57,355.49) .. (749.84,372.83) -- (677.33,307.13) -- cycle ; \draw  [color={rgb, 255:red, 126; green, 211; blue, 33 }  ,draw opacity=1 ][line width=1.5]  (749.35,240.88) .. controls (765.41,258.34) and (775.21,281.64) .. (775.18,307.23) .. controls (775.16,332.48) and (765.57,355.49) .. (749.84,372.83) ;  
%Curve Lines [id:da4836316055879939] 
\draw [color={rgb, 255:red, 0; green, 0; blue, 0 }  ,draw opacity=1 ][line width=0.75]    (604.29,242.02) .. controls (645.79,274.55) and (713.04,276.08) .. (749.35,240.88) ;
%Curve Lines [id:da27552280206279756] 
\draw [color={rgb, 255:red, 0; green, 0; blue, 0 }  ,draw opacity=1 ][line width=0.75]    (606.45,373.91) .. controls (640.57,335.88) and (713.55,337.73) .. (749.84,372.83) ;
%Curve Lines [id:da9863371559320575] 
\draw [color={rgb, 255:red, 245; green, 166; blue, 35 }  ,draw opacity=1 ][line width=1.5]    (607.09,375.25) .. controls (553.57,335.18) and (560.19,273.03) .. (604.29,242.02) ;
%Curve Lines [id:da7980061128078244] 
\draw [color={rgb, 255:red, 126; green, 211; blue, 33 }  ,draw opacity=1 ][line width=1.5]    (749.35,240.88) .. controls (803.73,279.14) and (788.95,351.49) .. (749.84,372.83) ;
%Curve Lines [id:da801002620610566] 
\draw [color={rgb, 255:red, 189; green, 16; blue, 224 }  ,draw opacity=1 ][line width=1.5]  [dash pattern={on 5.63pt off 4.5pt}]  (682.47,346.39) .. controls (672.28,333.15) and (668.2,286.78) .. (682.47,266.91) ;
%Curve Lines [id:da6152075064503373] 
\draw [color={rgb, 255:red, 189; green, 16; blue, 224 }  ,draw opacity=1 ][line width=1.5]    (682.47,266.91) .. controls (699.28,282.71) and (692.66,337.73) .. (682.47,346.39) ;
%Shape: Ellipse [id:dp3077062511388453] 
\draw   (251.74,308.13) .. controls (251.74,254.23) and (295.43,210.53) .. (349.33,210.53) .. controls (403.23,210.53) and (446.93,254.23) .. (446.93,308.13) .. controls (446.93,362.03) and (403.23,405.72) .. (349.33,405.72) .. controls (295.43,405.72) and (251.74,362.03) .. (251.74,308.13) -- cycle ;
%Shape: Arc [id:dp5931866487382411] 
\draw  [draw opacity=0][line width=1.5]  (279.09,376.25) .. controls (262,358.63) and (251.48,334.61) .. (251.48,308.13) .. controls (251.48,283.13) and (260.86,260.32) .. (276.29,243.02) -- (349.33,308.13) -- cycle ; \draw  [color={rgb, 255:red, 245; green, 166; blue, 35 }  ,draw opacity=1 ][line width=1.5]  (279.09,376.25) .. controls (262,358.63) and (251.48,334.61) .. (251.48,308.13) .. controls (251.48,283.13) and (260.86,260.32) .. (276.29,243.02) ;  
%Shape: Arc [id:dp18087195946135592] 
\draw  [draw opacity=0][line width=1.5]  (421.35,241.88) .. controls (437.41,259.34) and (447.21,282.64) .. (447.18,308.23) .. controls (447.16,333.48) and (437.57,356.49) .. (421.84,373.83) -- (349.33,308.13) -- cycle ; \draw  [color={rgb, 255:red, 126; green, 211; blue, 33 }  ,draw opacity=1 ][line width=1.5]  (421.35,241.88) .. controls (437.41,259.34) and (447.21,282.64) .. (447.18,308.23) .. controls (447.16,333.48) and (437.57,356.49) .. (421.84,373.83) ;  
%Curve Lines [id:da6203218881778263] 
\draw [color={rgb, 255:red, 245; green, 166; blue, 35 }  ,draw opacity=1 ][line width=1.5]    (276.29,243.02) .. controls (312.76,266.08) and (321.76,266.08) .. (349.76,267.08) ;
%Curve Lines [id:da022582853808335246] 
\draw [color={rgb, 255:red, 126; green, 211; blue, 33 }  ,draw opacity=1 ][line width=1.5]    (349.76,267.08) .. controls (373.76,270.08) and (397.76,261.08) .. (421.35,241.88) ;
%Straight Lines [id:da733266845690614] 
\draw [color={rgb, 255:red, 189; green, 16; blue, 224 }  ,draw opacity=1 ][line width=1.5]  [dash pattern={on 5.63pt off 4.5pt}]  (349.76,267.08) -- (349.76,349.89) ;
%Curve Lines [id:da17937535960533846] 
\draw [color={rgb, 255:red, 126; green, 211; blue, 33 }  ,draw opacity=1 ][line width=1.5]    (422.95,373.08) .. controls (386.2,350.45) and (377.2,350.56) .. (349.19,349.89) ;
%Curve Lines [id:da5817900908524667] 
\draw [color={rgb, 255:red, 245; green, 166; blue, 35 }  ,draw opacity=1 ][line width=1.5]    (349.19,349.89) .. controls (325.16,347.18) and (301.27,356.46) .. (277.91,375.94) ;

% Text Node
\draw (218.78,294.76) node [anchor=north west][inner sep=0.75pt]    {$\textcolor[rgb]{0.96,0.65,0.14}{A}$};
% Text Node
\draw (462.61,294.27) node [anchor=north west][inner sep=0.75pt]    {$\textcolor[rgb]{0.49,0.83,0.13}{B}$};
% Text Node
\draw (526,291.4) node [anchor=north west][inner sep=0.75pt]    {$\textcolor[rgb]{0.96,0.65,0.14}{AA^{*}}$};
% Text Node
\draw (799,292.4) node [anchor=north west][inner sep=0.75pt]    {$\textcolor[rgb]{0.49,0.83,0.13}{BB^{*}}$};
% Text Node
\draw (702,295.4) node [anchor=north west][inner sep=0.75pt]    {$\textcolor[rgb]{0.74,0.06,0.88}{S}\textcolor[rgb]{0.74,0.06,0.88}{_{R}}$};
% Text Node
\draw (362,295.4) node [anchor=north west][inner sep=0.75pt]    {$\textcolor[rgb]{0.74,0.06,0.88}{E}\textcolor[rgb]{0.74,0.06,0.88}{_{W}}$};
% Text Node
\draw (372.61,233.27) node [anchor=north west][inner sep=0.75pt]  [color={rgb, 255:red, 126; green, 211; blue, 33 }  ,opacity=1 ]  {$\textcolor[rgb]{0.49,0.83,0.13}{\overline{\textcolor[rgb]{0.49,0.83,0.13}{B}}}$};
% Text Node
\draw (371.61,372.27) node [anchor=north west][inner sep=0.75pt]    {$\textcolor[rgb]{0.49,0.83,0.13}{\overline{B}}$};
% Text Node
\draw (314.61,233.27) node [anchor=north west][inner sep=0.75pt]    {$\textcolor[rgb]{0.96,0.65,0.14}{\overline{\textcolor[rgb]{0.96,0.65,0.14}{A}}}$};
% Text Node
\draw (311.61,372.27) node [anchor=north west][inner sep=0.75pt]    {$\textcolor[rgb]{0.96,0.65,0.14}{\overline{A}}$};

\end{tikzpicture}

\par\end{centering}
\caption{\label{fig:purification} The bulk dual of the EoP ($E_W$ in left panel) and the bulk dual of the
reflected entropy ($S_R$ in right panel).}
\end{figure}

The key idea of the entanglement of purification (EoP) is to purify
a mixed state $\rho_{AB}$ by introducing an auxiliary system $\bar{A}\bar{B}$.
The EoP is then defined as the minimum von Neumann entropy between the purified state $A\bar{A}$ and
$B\bar{B}$  over
all possible purifications: $E_{p}=\mathrm{min}_{\bar{A}\bar{B}}S_{\mathrm{vN}}\left(A\bar{A}:B\bar{B}\right)$.
The holographic dual of  EoP is supposed to be a 
geodesic in the bulk of AdS$_{3}$, known as the entanglement
wedge cross-section (EWCS or $E_{W}$), as shown by the left panel  of Figure \ref{fig:purification}.  
However, optimization over purifications is, in practice, not workable
in CFTs. As a compromise, canonical purification is
often employed, which involves directly duplicating the mixed state
$AB$ into a state denoted $AA^{*}BB^{*}$. The resulting entanglement
entropy, called the reflected entropy, is given by $S_{R}\left(A:B\right)=-\mathrm{Tr}\rho_{AA^{*}}\log\rho_{AA^{*}}$.
In holographic interpretation, it has been shown that $S_{R}\left(A:B\right)=2E_{W}\left(A:B\right)$.
The bulk dual of the reflected entropy is a closed curve comprising two identical geodesics, 
as illustrated by the right panel of Figure \ref{fig:purification}.

To capture the amount of entanglement between two disjoint subsystems $A$ and $B$,  one must remove the over-counted degrees of freedom, denoted by $D_{\text{oc}}$, which arise from entanglement between $AB$ and its complement $(AB)^c$. For example, the EoP replaces $(AB)^c$ with auxiliary systems $\bar{A}\bar{B}$; the minimization procedure ensures that $D_{\text{oc}}$ vanishes. The reflected entropy replaces $(AB)^c$ with $A^* B^*$, a copy of the original system $AB$, which also ensures the vanishing of $D_{\text{oc}}$ but double-counts the entanglement between $A$ and $B$ due to identical entanglement between $A^*$ and $B^*$. Furthermore, in CFT, purifications involving auxiliary systems dramatically alter the path integral. Similar alterations occur in the partial transpose used to define entanglement negativity. In any case, the purpose is to remove the effects caused by  $(AB)^c$ with a controlled manner. 
To this end, why not seek  a covariant approach to remove  $(AB)^c$  within the path integral, thereby eliminating $D_{\text{oc}}$?

\begin{figure}[h]
\begin{centering}

\tikzset{every picture/.style={line width=0.75pt}} %set default line width to 0.75pt        

\begin{tikzpicture}[x=0.75pt,y=0.75pt,yscale=-1,xscale=1]
%uncomment if require: \path (0,502); %set diagram left start at 0, and has height of 502

%Shape: Rectangle [id:dp8803082999798688] 
\draw  [color={rgb, 255:red, 0; green, 0; blue, 0 }  ,draw opacity=0 ][fill={rgb, 255:red, 74; green, 144; blue, 226 }  ,fill opacity=0.3 ] (245.49,146.38) -- (489.65,146.38) -- (489.65,383.69) -- (245.49,383.69) -- cycle ;
%Straight Lines [id:da6115139094920372] 
\draw [color={rgb, 255:red, 245; green, 166; blue, 35 }  ,draw opacity=1 ][line width=2.25]    (244.36,271.2) -- (355.96,271.2) ;
%Straight Lines [id:da4485215018607842] 
\draw [color={rgb, 255:red, 126; green, 211; blue, 33 }  ,draw opacity=1 ][line width=2.25]    (369.15,271.2) -- (489.65,271.2) ;
%Shape: Ellipse [id:dp5681963604281945] 
\draw  [fill={rgb, 255:red, 255; green, 255; blue, 255 }  ,fill opacity=1 ] (355.96,271.2) .. controls (355.96,267.56) and (358.91,264.61) .. (362.55,264.61) .. controls (366.2,264.61) and (369.15,267.56) .. (369.15,271.2) .. controls (369.15,274.85) and (366.2,277.8) .. (362.55,277.8) .. controls (358.91,277.8) and (355.96,274.85) .. (355.96,271.2) -- cycle ;
%Shape: Rectangle [id:dp3443295518675489] 
\draw  [color={rgb, 255:red, 0; green, 0; blue, 0 }  ,draw opacity=0 ][fill={rgb, 255:red, 74; green, 144; blue, 226 }  ,fill opacity=0.3 ] (601.49,145.38) -- (845.65,145.38) -- (845.65,382.69) -- (601.49,382.69) -- cycle ;
%Straight Lines [id:da13797472883471484] 
\draw [color={rgb, 255:red, 245; green, 166; blue, 35 }  ,draw opacity=1 ][line width=2.25]    (600.36,270.2) -- (711.96,270.2) ;
%Straight Lines [id:da8689944320392613] 
\draw [color={rgb, 255:red, 126; green, 211; blue, 33 }  ,draw opacity=1 ][line width=2.25]    (725.15,270.2) -- (845.65,270.2) ;
%Shape: Ellipse [id:dp7058253154105891] 
\draw  [fill={rgb, 255:red, 255; green, 255; blue, 255 }  ,fill opacity=1 ][dash pattern={on 4.5pt off 4.5pt}] (684.59,270.2) .. controls (684.59,247.8) and (702.75,229.64) .. (725.15,229.64) .. controls (747.55,229.64) and (765.71,247.8) .. (765.71,270.2) .. controls (765.71,292.6) and (747.55,310.76) .. (725.15,310.76) .. controls (702.75,310.76) and (684.59,292.6) .. (684.59,270.2) -- cycle ;
%Straight Lines [id:da9726301787594505] 
\draw    (265,373) -- (265,335) ;
\draw [shift={(265,333)}, rotate = 90] [color={rgb, 255:red, 0; green, 0; blue, 0 }  ][line width=0.75]    (10.93,-3.29) .. controls (6.95,-1.4) and (3.31,-0.3) .. (0,0) .. controls (3.31,0.3) and (6.95,1.4) .. (10.93,3.29)   ;
%Straight Lines [id:da14179960466049035] 
\draw    (265,373) -- (303,373) ;
\draw [shift={(305,373)}, rotate = 180] [color={rgb, 255:red, 0; green, 0; blue, 0 }  ][line width=0.75]    (10.93,-3.29) .. controls (6.95,-1.4) and (3.31,-0.3) .. (0,0) .. controls (3.31,0.3) and (6.95,1.4) .. (10.93,3.29)   ;
%Straight Lines [id:da07449877757149115] 
\draw    (617,371) -- (617,333) ;
\draw [shift={(617,331)}, rotate = 90] [color={rgb, 255:red, 0; green, 0; blue, 0 }  ][line width=0.75]    (10.93,-3.29) .. controls (6.95,-1.4) and (3.31,-0.3) .. (0,0) .. controls (3.31,0.3) and (6.95,1.4) .. (10.93,3.29)   ;
%Straight Lines [id:da4814031649911854] 
\draw    (617,371) -- (655,371) ;
\draw [shift={(657,371)}, rotate = 180] [color={rgb, 255:red, 0; green, 0; blue, 0 }  ][line width=0.75]    (10.93,-3.29) .. controls (6.95,-1.4) and (3.31,-0.3) .. (0,0) .. controls (3.31,0.3) and (6.95,1.4) .. (10.93,3.29)   ;

% Text Node
\draw (299.8,244.38) node [anchor=north west][inner sep=0.75pt]  [color={rgb, 255:red, 208; green, 2; blue, 27 }  ,opacity=1 ]  {$\textcolor[rgb]{0.96,0.65,0.14}{A}$};
% Text Node
\draw (418.01,244.38) node [anchor=north west][inner sep=0.75pt]    {$\textcolor[rgb]{0.49,0.83,0.13}{B}$};
% Text Node
\draw (444.33,284.52) node [anchor=north west][inner sep=0.75pt]    {$\tau =0$};
% Text Node
\draw (358.27,283.82) node [anchor=north west][inner sep=0.75pt]    {$\epsilon $};
% Text Node
\draw (648.8,244.38) node [anchor=north west][inner sep=0.75pt]  [color={rgb, 255:red, 208; green, 2; blue, 27 }  ,opacity=1 ]  {$\textcolor[rgb]{0.96,0.65,0.14}{A}$};
% Text Node
\draw (791.01,244.38) node [anchor=north west][inner sep=0.75pt]    {$\textcolor[rgb]{0.49,0.83,0.13}{B}$};
% Text Node
\draw (799.33,283.52) node [anchor=north west][inner sep=0.75pt]    {$\tau =0$};
% Text Node
\draw (307,356.4) node [anchor=north west][inner sep=0.75pt]    {$x$};
% Text Node
\draw (267,316.4) node [anchor=north west][inner sep=0.75pt]    {$\tau $};
% Text Node
\draw (659,354.4) node [anchor=north west][inner sep=0.75pt]    {$x$};
% Text Node
\draw (619,314.4) node [anchor=north west][inner sep=0.75pt]    {$\tau $};

\end{tikzpicture}

\par\end{centering}
\caption{The subtraction method could be understood as introducing a finite
regulator. Left panel: Two subsystems with an infinitesimal regulator.
Right panel: Two subsystems with a finite regulator.\label{fig:subtract region}}
\end{figure}
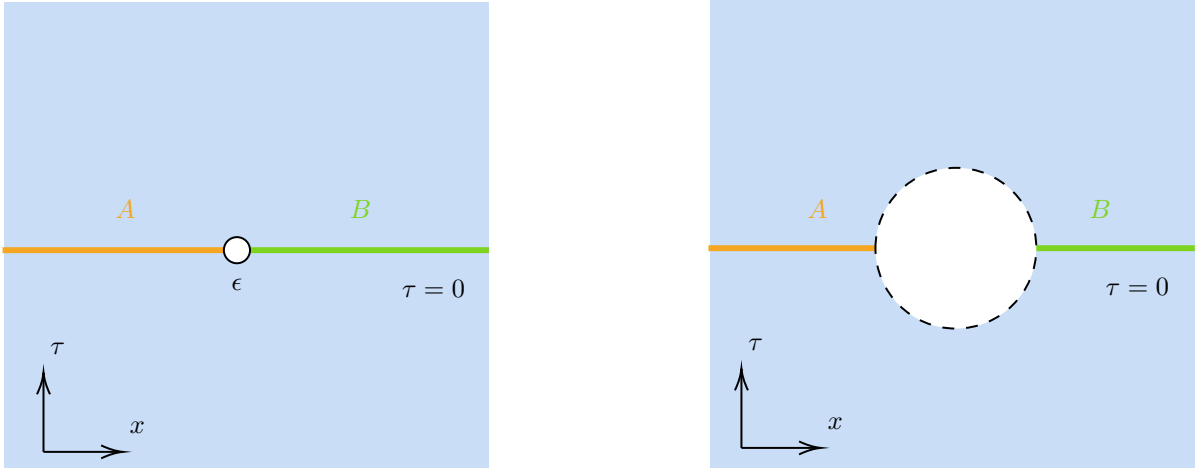

\begin{figure}[b]

\tikzset{every picture/.style={line width=0.75pt}} %set default line width to 0.75pt        

\begin{tikzpicture}[x=0.75pt,y=0.75pt,yscale=-1,xscale=1]
%uncomment if require: \path (0,618); %set diagram left start at 0, and has height of 618

%Shape: Rectangle [id:dp985427242444793] 
\draw  [color={rgb, 255:red, 0; green, 0; blue, 0 }  ,draw opacity=0 ][fill={rgb, 255:red, 74; green, 144; blue, 226 }  ,fill opacity=0.3 ] (473.49,196.38) -- (717.65,196.38) -- (717.65,433.69) -- (473.49,433.69) -- cycle ;
%Shape: Rectangle [id:dp6042754230130538] 
\draw  [color={rgb, 255:red, 0; green, 0; blue, 0 }  ,draw opacity=0 ][fill={rgb, 255:red, 74; green, 144; blue, 226 }  ,fill opacity=0.3 ] (117.49,197.38) -- (361.65,197.38) -- (361.65,434.69) -- (117.49,434.69) -- cycle ;
%Straight Lines [id:da6179170738452158] 
\draw [color={rgb, 255:red, 126; green, 211; blue, 33 }  ,draw opacity=1 ][line width=3]    (260,297) -- (320,297) ;
%Straight Lines [id:da278767186737771] 
\draw [color={rgb, 255:red, 245; green, 166; blue, 35 }  ,draw opacity=1 ][line width=3]    (140,297) -- (195.5,297) -- (200,297) ;
%Straight Lines [id:da3518953022322977] 
\draw [color={rgb, 255:red, 126; green, 211; blue, 33 }  ,draw opacity=1 ][line width=3]    (620,277) -- (680,257) ;
%Straight Lines [id:da4491102464468042] 
\draw [color={rgb, 255:red, 245; green, 166; blue, 35 }  ,draw opacity=1 ][line width=3]    (520,257) -- (580,277) ;

% Text Node
\draw (530,228.4) node [anchor=north west][inner sep=0.75pt]  [font=\large,color={rgb, 255:red, 245; green, 166; blue, 35 }  ,opacity=1 ]  {$A$};
% Text Node
\draw (641.22,230.37) node [anchor=north west][inner sep=0.75pt]  [font=\Large,color={rgb, 255:red, 126; green, 211; blue, 33 }  ,opacity=1 ]  {$B$};
% Text Node
\draw (502,260.4) node [anchor=north west][inner sep=0.75pt]    {$z_{1}$};
% Text Node
\draw (562,280.4) node [anchor=north west][inner sep=0.75pt]    {$z_{2}$};
% Text Node
\draw (622,280.4) node [anchor=north west][inner sep=0.75pt]    {$z_{3}$};
% Text Node
\draw (682,260.4) node [anchor=north west][inner sep=0.75pt]    {$z_{4}$};
% Text Node
\draw (142,300.4) node [anchor=north west][inner sep=0.75pt]    {$x_{1}$};
% Text Node
\draw (202,300.4) node [anchor=north west][inner sep=0.75pt]    {$x_{2}$};
% Text Node
\draw (262,300.4) node [anchor=north west][inner sep=0.75pt]    {$x_{3}$};
% Text Node
\draw (322,300.4) node [anchor=north west][inner sep=0.75pt]    {$x_{4}$};
% Text Node
\draw (162,260.4) node [anchor=north west][inner sep=0.75pt]  [font=\large,color={rgb, 255:red, 245; green, 166; blue, 35 }  ,opacity=1 ]  {$A$};
% Text Node
\draw (292.22,256.37) node [anchor=north west][inner sep=0.75pt]  [font=\Large,color={rgb, 255:red, 126; green, 211; blue, 33 }  ,opacity=1 ]  {$B$};

\end{tikzpicture}

\caption{Two different configurations of mixed states in CFT living on the
complex plane, where the blue shaded regions represent the Euclidean
path integral. Left panel: At $\tau=0$, two subsystems $A$ and $B$
on the $x$-axis in a mixed state $\rho_{AB}$. Right panel: Two generic
subsystems that are determined by four ending points $z_i$ at different
time.\label{fig:setup}}

\end{figure}
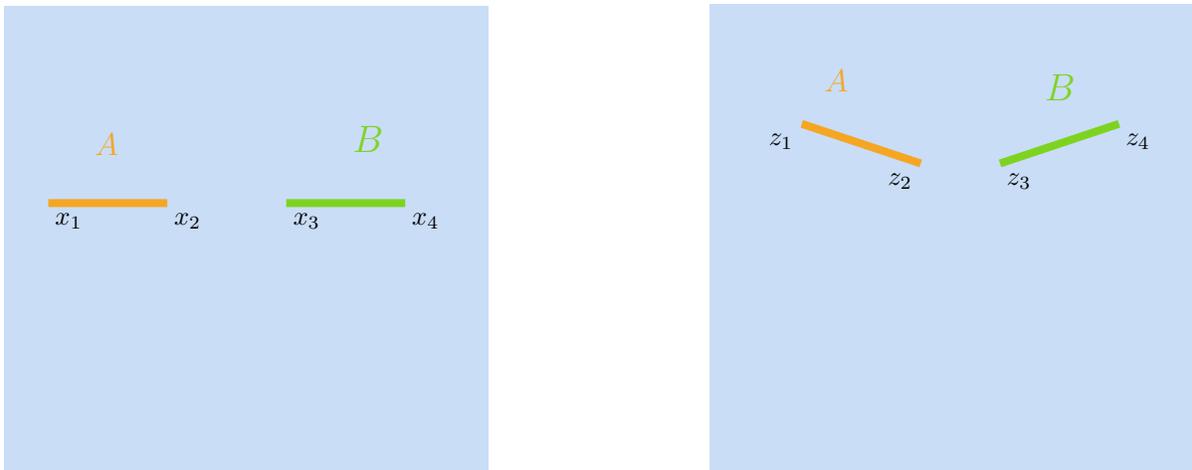

In \citep{Jiang:2024ijx}, we proposed an alternative approach, namely SUBTRACTION,  to
purification in CFT$_{2}$. Instead of purifying the mixed state by
adding auxiliary systems, our method subtracts the undetectable regions
of the system, as illustrated in Figure \ref{fig:subtract region}.
The remaining regions, which are disjoint but mutually complementary,
form a pure entangled state $\psi_{AB}$, satisfying $A^{C}=B$. 
This process is equivalent to enlarging the cutoff regulators, effectively
situating the boundary CFT (BCFT) on the edge of the excluded region
(dashed circle). Using the replica trick \citep{Calabrese:2004eu,Calabrese:2009qy}
and the annulus CFT techniques \citep{Cardy:1989ir,Cardy:2004hm,Cardy:2016fqc},
we computed the von Neumann entropy $S_{\mathrm{vN}}\left(A:B\right)$
for two disjoint intervals $A$ and $B$ living at the time $t=0$,
as illustrated in Figure \ref{fig:setup}. Notably, it was found that
\begin{equation}
S_{\mathrm{vN}}\left(A:B\right)=E_{W}\left(A:B\right)
\end{equation}
for a \emph{static} time slice of bulk AdS$_{3}$. This proposal is
independent of purifications and four-point conformal block data,
making it very easy to compute. In this paper,
our aim is to extend the subtraction method to the covariant case
and compute the corresponding entropy $S_{\mathrm{vN}}\left(A:B\right)$
for two generic intervals $A$ and $B$, as illustrated in Figure
\ref{fig:setup}.
Beyond the advantage of computational simplicity, 
this approach offers two additional benefits: 
\begin{itemize}
\item First, the holographic dual of reflected entropy is a closed curve formed by two 
identical geodesics. By contrast, as we will demonstrate, the holographic dual of the SUBTRACTION-defined
$S_{\mathrm{vN}}\left(A:B\right)$ precisely equals the EWCS.
This critical property enables the differentiation of  $S_{\mathrm{vN}}\left(A:B\right)$
at its endpoints and the implementation of a coincident limit---procedures 
that allow for a straightforward derivation of the Einstein equation from
$S_{\mathrm{vN}}\left(A:B\right)$, as shown in \cite{Jiang:2024xcy}. 
\item Second, if we take the annulus---the core product of the SUBTRACTION approach---as the starting point, a striking generalization emerges: this framework can be easily extended to higher dimensions, specifically to the solid torus \(D^{D-1}\times S^1\). As shown in \cite{Jiang:2025jnk}, this extension facilitates the straightforward calculation of entanglement entropy of CFT$_D$ for all $D$.
\end{itemize}

%\textcolor{red}{As we will see, the holographic dual of such defined $S_{\mathrm{vN}}\left(A:B\right)$
%precisely equals EWCS. Since the holographic dual of the reflected entropy is a closed curve formed
%by two identical geodesics, 
%In contrast to the closed curve}

In the subtraction method, enlarging the regulator is reminiscent
of the renormalization group (RG) flow. Using the Callan-Symanzik
RG equation, we can derive an RG flow-induced entanglement entropy,
which turns out to be exactly $S_{\mathrm{vN}}(A:B)$ in CFT$_2$. Furthermore,
since the RG equation is universal, we can apply it to higher even dimensional
cases to extract some finite quantities. As an example, for a particular configuration in 
$\mathrm{CFT}_{4}$,
we calculate the RG flow induced entanglement
entropy and find that it agrees
with the entanglement wedge cross-section in the $\mathrm{AdS}_{5}$
bulk.

The remainder of this paper is structured as follows:  In Section \ref{sec:rigor}, we provide a rigorous calculation of covariant mixed state $S_{\mathrm{vN}}\left(A:B\right)$ using the replica trick in the
subtraction method. We clarify that there are two gauge parameters in the entanglement
entropy $S_{\mathrm{vN}}(A:B)$.
In Section \ref{sec:Holography},
we discuss the holographic dual of $S_{\mathrm{vN}}(A:B)$. 
In Section \ref{sec:Renormalization-group-flow},
we define the RG flow induced entanglement entropy through the Callan-Symanzik
RG equation. We then apply it to  CFT$_{2}$ and a particular configuration in CFT$_{4}$, respectively.
Section \ref{sec:Conclusion} contains the conclusion.

\section{Covariant entanglement entropy for mixed states\label{sec:rigor}}
In previous work \citep{Jiang:2024ijx}, we introduced the SUBTRACTION approach, as an alternative
to purification in CFT. The computed quantity, $S_{\mathrm{vN}}\left(A:B\right)$,
is the entropy of entanglement between two disjoint intervals $A$
and $B$ in a mixed state $\rho_{AB}$ on a time slice. To complete the story, it is of importance 
to include the time dependence in the expression of $S_{\mathrm{vN}}\left(A:B\right)$. 
%We will achieve
%this by using  the method of  boosting the entangling region on the complex plane.
%In this section, we verify the above discussion by rigorously computing
%the von Neumann entropy in the same setup. 

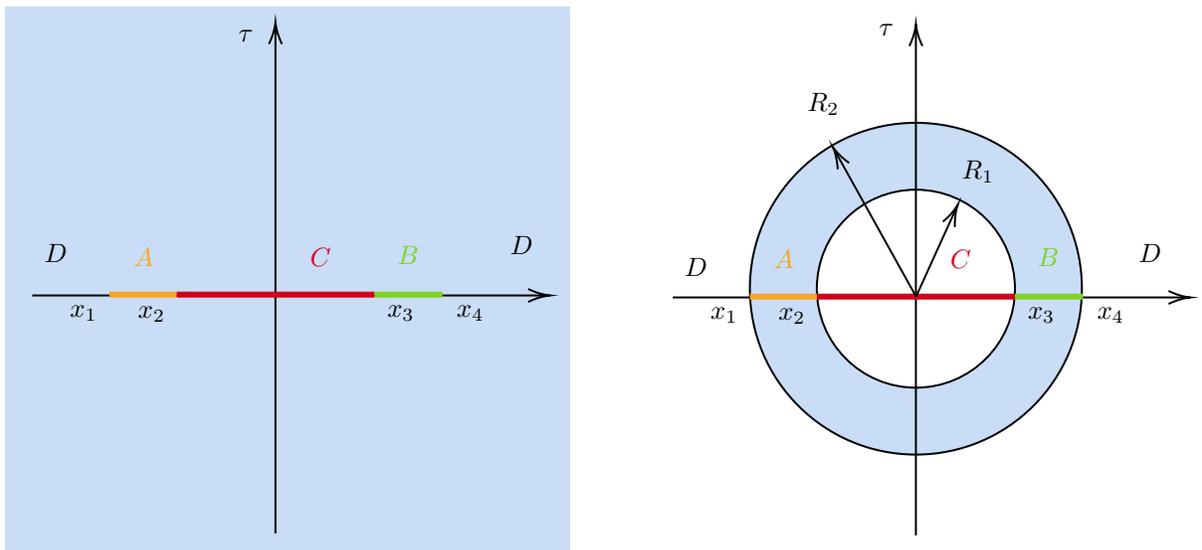
\begin{figure}[h]
\begin{centering}

\tikzset{every picture/.style={line width=0.75pt}} %set default line width to 0.75pt        

\begin{tikzpicture}[x=0.75pt,y=0.75pt,yscale=-1,xscale=1]

%Shape: Circle [id:dp4757863490578256] 
\draw  [fill={rgb, 255:red, 74; green, 144; blue, 226 }  ,fill opacity=0.3 ] (419.27,148.5) .. controls (419.27,102.26) and (456.76,64.77) .. (503,64.77) .. controls (549.24,64.77) and (586.73,102.26) .. (586.73,148.5) .. controls (586.73,194.74) and (549.24,232.23) .. (503,232.23) .. controls (456.76,232.23) and (419.27,194.74) .. (419.27,148.5) -- cycle ;
%Shape: Circle [id:dp9149979206579004] 
\draw  [fill={rgb, 255:red, 255; green, 255; blue, 255 }  ,fill opacity=1 ] (453.04,148.5) .. controls (453.04,120.91) and (475.41,98.54) .. (503,98.54) .. controls (530.59,98.54) and (552.96,120.91) .. (552.96,148.5) .. controls (552.96,176.09) and (530.59,198.46) .. (503,198.46) .. controls (475.41,198.46) and (453.04,176.09) .. (453.04,148.5) -- cycle ;
%Shape: Rectangle [id:dp8755803041956567] 
\draw  [color={rgb, 255:red, 0; green, 0; blue, 0 }  ,draw opacity=0 ][fill={rgb, 255:red, 74; green, 144; blue, 226 }  ,fill opacity=0.3 ] (43.68,6.38) -- (328.36,6.38) -- (328.36,283.08) -- (43.68,283.08) -- cycle ;
%Straight Lines [id:da21064689753935828] 
\draw    (57.36,151.92) -- (316.36,151.92) ;
\draw [shift={(318.36,151.92)}, rotate = 180] [color={rgb, 255:red, 0; green, 0; blue, 0 }  ][line width=0.75]    (10.93,-3.29) .. controls (6.95,-1.4) and (3.31,-0.3) .. (0,0) .. controls (3.31,0.3) and (6.95,1.4) .. (10.93,3.29)   ;
%Straight Lines [id:da4376222667098817] 
\draw    (180,272) -- (180,16.08) ;
\draw [shift={(180,14.08)}, rotate = 90] [color={rgb, 255:red, 0; green, 0; blue, 0 }  ][line width=0.75]    (10.93,-3.29) .. controls (6.95,-1.4) and (3.31,-0.3) .. (0,0) .. controls (3.31,0.3) and (6.95,1.4) .. (10.93,3.29)   ;
%Straight Lines [id:da2306762023706206] 
\draw [color={rgb, 255:red, 245; green, 166; blue, 35 }  ,draw opacity=1 ][line width=2.25]    (96.27,151.5) -- (130.24,151.5) ;
%Straight Lines [id:da7138417931940471] 
\draw [color={rgb, 255:red, 208; green, 2; blue, 27 }  ,draw opacity=1 ][line width=2.25]    (130.24,151.5) -- (229.96,151.5) ;
%Straight Lines [id:da8504550848245445] 
\draw [color={rgb, 255:red, 126; green, 211; blue, 33 }  ,draw opacity=1 ][line width=2.25]    (229.96,151.5) -- (264.24,151.5) ;
%Straight Lines [id:da226659218342947] 
\draw    (380.36,152.92) -- (639.36,152.92) ;
\draw [shift={(641.36,152.92)}, rotate = 180] [color={rgb, 255:red, 0; green, 0; blue, 0 }  ][line width=0.75]    (10.93,-3.29) .. controls (6.95,-1.4) and (3.31,-0.3) .. (0,0) .. controls (3.31,0.3) and (6.95,1.4) .. (10.93,3.29)   ;
%Straight Lines [id:da18635246717440368] 
\draw    (503,273) -- (503,17.08) ;
\draw [shift={(503,15.08)}, rotate = 90] [color={rgb, 255:red, 0; green, 0; blue, 0 }  ][line width=0.75]    (10.93,-3.29) .. controls (6.95,-1.4) and (3.31,-0.3) .. (0,0) .. controls (3.31,0.3) and (6.95,1.4) .. (10.93,3.29)   ;
%Straight Lines [id:da30572025288933924] 
\draw [color={rgb, 255:red, 245; green, 166; blue, 35 }  ,draw opacity=1 ][line width=2.25]    (419.27,152.5) -- (453.24,152.5) ;
%Straight Lines [id:da12004333701859493] 
\draw [color={rgb, 255:red, 208; green, 2; blue, 27 }  ,draw opacity=1 ][line width=2.25]    (453.24,152.5) -- (552.96,152.5) ;
%Straight Lines [id:da6297380858126034] 
\draw [color={rgb, 255:red, 126; green, 211; blue, 33 }  ,draw opacity=1 ][line width=2.25]    (552.96,152.5) -- (587.24,152.5) ;
%Straight Lines [id:da9771848049226468] 
\draw    (503.1,152.5) -- (462.13,78.83) ;
\draw [shift={(461.16,77.08)}, rotate = 60.92] [color={rgb, 255:red, 0; green, 0; blue, 0 }  ][line width=0.75]    (10.93,-3.29) .. controls (6.95,-1.4) and (3.31,-0.3) .. (0,0) .. controls (3.31,0.3) and (6.95,1.4) .. (10.93,3.29)   ;
%Straight Lines [id:da27346564695425624] 
\draw    (503.1,152.5) -- (523.33,107.9) ;
\draw [shift={(524.16,106.08)}, rotate = 114.4] [color={rgb, 255:red, 0; green, 0; blue, 0 }  ][line width=0.75]    (10.93,-3.29) .. controls (6.95,-1.4) and (3.31,-0.3) .. (0,0) .. controls (3.31,0.3) and (6.95,1.4) .. (10.93,3.29)   ;

% Text Node
\draw (109.26,155.9) node [anchor=north west][inner sep=0.75pt]    {$x_{2}$};
% Text Node
\draw (235,155.32) node [anchor=north west][inner sep=0.75pt]    {$x_{3}$};
% Text Node
\draw (270,155.32) node [anchor=north west][inner sep=0.75pt]    {$x_{4}$};
% Text Node
\draw (75,155.32) node [anchor=north west][inner sep=0.75pt]    {$x_{1}$};
% Text Node
\draw (107,126.4) node [anchor=north west][inner sep=0.75pt]    {$\textcolor[rgb]{0.96,0.65,0.14}{A}$};
% Text Node
\draw (240,125.4) node [anchor=north west][inner sep=0.75pt]    {$\textcolor[rgb]{0.49,0.83,0.13}{B}$};
% Text Node
\draw (196,126.4) node [anchor=north west][inner sep=0.75pt]    {$\textcolor[rgb]{0.82,0.01,0.11}{C}$};
% Text Node
\draw (432.26,156.9) node [anchor=north west][inner sep=0.75pt]    {$x_{2}$};
% Text Node
\draw (558,156.32) node [anchor=north west][inner sep=0.75pt]    {$x_{3}$};
% Text Node
\draw (593,156.32) node [anchor=north west][inner sep=0.75pt]    {$x_{4}$};
% Text Node
\draw (398,156.32) node [anchor=north west][inner sep=0.75pt]    {$x_{1}$};
% Text Node
\draw (430,127.4) node [anchor=north west][inner sep=0.75pt]    {$\textcolor[rgb]{0.96,0.65,0.14}{A}$};
% Text Node
\draw (563,126.4) node [anchor=north west][inner sep=0.75pt]    {$\textcolor[rgb]{0.49,0.83,0.13}{B}$};
% Text Node
\draw (519,127.4) node [anchor=north west][inner sep=0.75pt]    {$\textcolor[rgb]{0.82,0.01,0.11}{C}$};
% Text Node
\draw (446.8,48.4) node [anchor=north west][inner sep=0.75pt]    {$R_{2}$};
% Text Node
\draw (524.8,82.4) node [anchor=north west][inner sep=0.75pt]    {$R_{1}$};
% Text Node
\draw (160,16.4) node [anchor=north west][inner sep=0.75pt]    {$\tau $};
% Text Node
\draw (483,13.4) node [anchor=north west][inner sep=0.75pt]    {$\tau $};
% Text Node
\draw (385,131.4) node [anchor=north west][inner sep=0.75pt]    {$D$};
% Text Node
\draw (62,124.4) node [anchor=north west][inner sep=0.75pt]    {$D$};
% Text Node
\draw (297,120.4) node [anchor=north west][inner sep=0.75pt]    {$D$};
% Text Node
\draw (614,124.4) node [anchor=north west][inner sep=0.75pt]    {$D$};

\end{tikzpicture}
\par\end{centering}
\caption{Static symmetric configuration.  Left panel: At $\tau=0$, two subsystems $A$ and $B$ on the $x$-axis
are in a mixed state $\rho_{AB}$, separated by segments $C$ and $D$. Right panel: After introducing two
finite regulators (subtracting $C$ and $D$ with two discs), we
obtain an annulus, in which two subsystems $A$ and $B$ are now in
a pure entangled state $\psi_{AB}$. \label{fig:alternative puri}}
\end{figure}

\subsection{Symmetric configuration and Replica trick}

We begin by considering the symmetric case and will extend to  generic configurations using conformal transformations.
As explained in the introduction, in ref.  \citep{Jiang:2024ijx}, we considered two disjoint segments $A$ and $B$ on the $x-$axis at $\tau=0$, which are separated by undetectable segments $C$ and $D$, as illustrated in Figure \ref{fig:alternative puri}. Segments $C$ and $D$ are then removed  by two discs with conformal invariant boundary conditions, which encode the irrelevant information coming from $C$ and $D$.

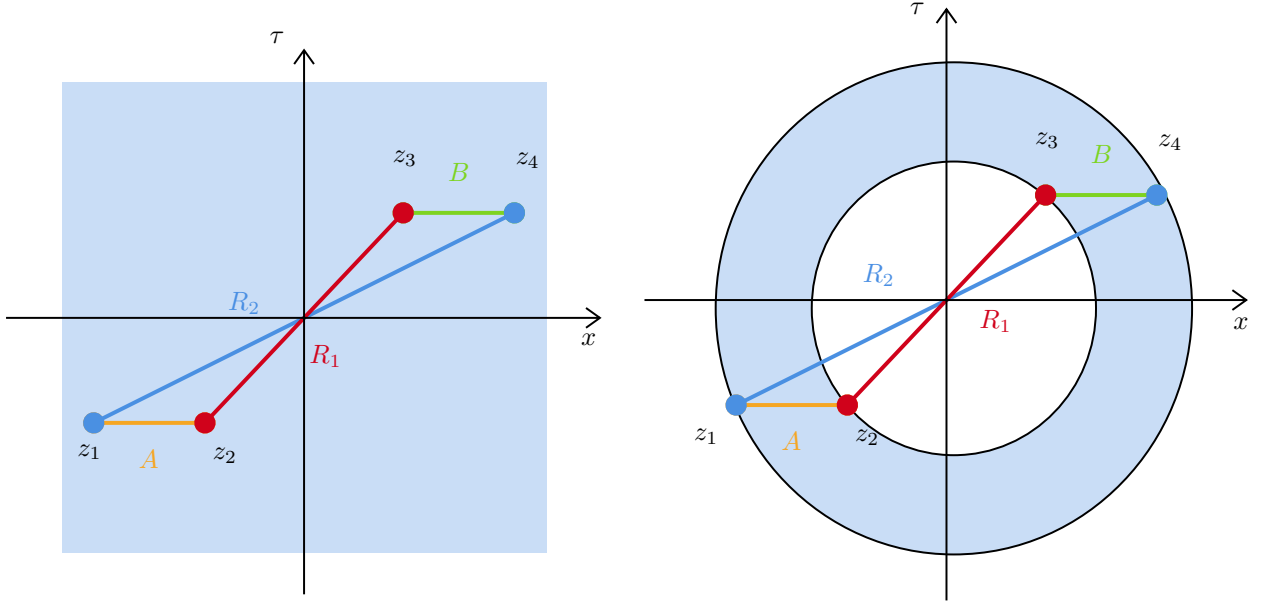
\begin{figure}[h]

\tikzset{every picture/.style={line width=0.75pt}} %set default line width to 0.75pt        

\begin{tikzpicture}[x=0.75pt,y=0.75pt,yscale=-1,xscale=1]
%uncomment if require: \path (0,419); %set diagram left start at 0, and has height of 419

%Shape: Rectangle [id:dp15235918958936012] 
\draw  [color={rgb, 255:red, 0; green, 0; blue, 0 }  ,draw opacity=0 ][fill={rgb, 255:red, 74; green, 144; blue, 226 }  ,fill opacity=0.3 ] (173.68,104.5) -- (417.84,104.5) -- (417.84,341.82) -- (173.68,341.82) -- cycle ;
%Shape: Ellipse [id:dp26883410353429693] 
\draw  [fill={rgb, 255:red, 74; green, 144; blue, 226 }  ,fill opacity=0.3 ] (503.36,218.36) .. controls (503.36,149.77) and (557.14,94.16) .. (623.49,94.16) .. controls (689.84,94.16) and (743.62,149.77) .. (743.62,218.36) .. controls (743.62,286.96) and (689.84,342.57) .. (623.49,342.57) .. controls (557.14,342.57) and (503.36,286.96) .. (503.36,218.36) -- cycle ;
%Shape: Ellipse [id:dp10544071562300639] 
\draw  [fill={rgb, 255:red, 255; green, 255; blue, 255 }  ,fill opacity=1 ] (551.81,218.36) .. controls (551.81,177.43) and (583.9,144.25) .. (623.49,144.25) .. controls (663.08,144.25) and (695.17,177.43) .. (695.17,218.36) .. controls (695.17,259.29) and (663.08,292.47) .. (623.49,292.47) .. controls (583.9,292.47) and (551.81,259.29) .. (551.81,218.36) -- cycle ;

%Straight Lines [id:da6263588659735757] 
\draw [color={rgb, 255:red, 245; green, 166; blue, 35 }  ,draw opacity=1 ][line width=1.5]    (513.68,267.16) -- (569.76,267.16) ;
\draw [shift={(569.76,267.16)}, rotate = 0] [color={rgb, 255:red, 245; green, 166; blue, 35 }  ,draw opacity=1 ][fill={rgb, 255:red, 245; green, 166; blue, 35 }  ,fill opacity=1 ][line width=1.5]      (0, 0) circle [x radius= 4.36, y radius= 4.36]   ;
\draw [shift={(513.68,267.16)}, rotate = 0] [color={rgb, 255:red, 245; green, 166; blue, 35 }  ,draw opacity=1 ][fill={rgb, 255:red, 245; green, 166; blue, 35 }  ,fill opacity=1 ][line width=1.5]      (0, 0) circle [x radius= 4.36, y radius= 4.36]   ;
%Straight Lines [id:da32396065084399317] 
\draw [color={rgb, 255:red, 126; green, 211; blue, 33 }  ,draw opacity=1 ][line width=1.5]    (669.76,161.16) -- (725.84,161.16) ;
\draw [shift={(725.84,161.16)}, rotate = 0] [color={rgb, 255:red, 126; green, 211; blue, 33 }  ,draw opacity=1 ][fill={rgb, 255:red, 126; green, 211; blue, 33 }  ,fill opacity=1 ][line width=1.5]      (0, 0) circle [x radius= 4.36, y radius= 4.36]   ;
\draw [shift={(669.76,161.16)}, rotate = 0] [color={rgb, 255:red, 126; green, 211; blue, 33 }  ,draw opacity=1 ][fill={rgb, 255:red, 126; green, 211; blue, 33 }  ,fill opacity=1 ][line width=1.5]      (0, 0) circle [x radius= 4.36, y radius= 4.36]   ;
%Straight Lines [id:da33176727888498303] 
\draw [color={rgb, 255:red, 74; green, 144; blue, 226 }  ,draw opacity=1 ][line width=1.5]    (513.68,267.16) -- (725.84,161.16) ;
\draw [shift={(725.84,161.16)}, rotate = 333.45] [color={rgb, 255:red, 74; green, 144; blue, 226 }  ,draw opacity=1 ][fill={rgb, 255:red, 74; green, 144; blue, 226 }  ,fill opacity=1 ][line width=1.5]      (0, 0) circle [x radius= 4.36, y radius= 4.36]   ;
\draw [shift={(513.68,267.16)}, rotate = 333.45] [color={rgb, 255:red, 74; green, 144; blue, 226 }  ,draw opacity=1 ][fill={rgb, 255:red, 74; green, 144; blue, 226 }  ,fill opacity=1 ][line width=1.5]      (0, 0) circle [x radius= 4.36, y radius= 4.36]   ;
%Straight Lines [id:da6553716220801873] 
\draw [color={rgb, 255:red, 208; green, 2; blue, 27 }  ,draw opacity=1 ][line width=1.5]    (569.76,267.16) -- (669.76,161.16) ;
\draw [shift={(669.76,161.16)}, rotate = 313.33] [color={rgb, 255:red, 208; green, 2; blue, 27 }  ,draw opacity=1 ][fill={rgb, 255:red, 208; green, 2; blue, 27 }  ,fill opacity=1 ][line width=1.5]      (0, 0) circle [x radius= 4.36, y radius= 4.36]   ;
\draw [shift={(569.76,267.16)}, rotate = 313.33] [color={rgb, 255:red, 208; green, 2; blue, 27 }  ,draw opacity=1 ][fill={rgb, 255:red, 208; green, 2; blue, 27 }  ,fill opacity=1 ][line width=1.5]      (0, 0) circle [x radius= 4.36, y radius= 4.36]   ;
%Straight Lines [id:da9978501854270239] 
\draw [color={rgb, 255:red, 245; green, 166; blue, 35 }  ,draw opacity=1 ][line width=1.5]    (189.68,276.16) -- (245.76,276.16) ;
\draw [shift={(245.76,276.16)}, rotate = 0] [color={rgb, 255:red, 245; green, 166; blue, 35 }  ,draw opacity=1 ][fill={rgb, 255:red, 245; green, 166; blue, 35 }  ,fill opacity=1 ][line width=1.5]      (0, 0) circle [x radius= 4.36, y radius= 4.36]   ;
\draw [shift={(189.68,276.16)}, rotate = 0] [color={rgb, 255:red, 245; green, 166; blue, 35 }  ,draw opacity=1 ][fill={rgb, 255:red, 245; green, 166; blue, 35 }  ,fill opacity=1 ][line width=1.5]      (0, 0) circle [x radius= 4.36, y radius= 4.36]   ;
%Straight Lines [id:da4974785037040834] 
\draw [color={rgb, 255:red, 126; green, 211; blue, 33 }  ,draw opacity=1 ][line width=1.5]    (345.76,170.16) -- (401.84,170.16) ;
\draw [shift={(401.84,170.16)}, rotate = 0] [color={rgb, 255:red, 126; green, 211; blue, 33 }  ,draw opacity=1 ][fill={rgb, 255:red, 126; green, 211; blue, 33 }  ,fill opacity=1 ][line width=1.5]      (0, 0) circle [x radius= 4.36, y radius= 4.36]   ;
\draw [shift={(345.76,170.16)}, rotate = 0] [color={rgb, 255:red, 126; green, 211; blue, 33 }  ,draw opacity=1 ][fill={rgb, 255:red, 126; green, 211; blue, 33 }  ,fill opacity=1 ][line width=1.5]      (0, 0) circle [x radius= 4.36, y radius= 4.36]   ;
%Straight Lines [id:da28578505339221594] 
\draw [color={rgb, 255:red, 74; green, 144; blue, 226 }  ,draw opacity=1 ][line width=1.5]    (189.68,276.16) -- (401.84,170.16) ;
\draw [shift={(401.84,170.16)}, rotate = 333.45] [color={rgb, 255:red, 74; green, 144; blue, 226 }  ,draw opacity=1 ][fill={rgb, 255:red, 74; green, 144; blue, 226 }  ,fill opacity=1 ][line width=1.5]      (0, 0) circle [x radius= 4.36, y radius= 4.36]   ;
\draw [shift={(189.68,276.16)}, rotate = 333.45] [color={rgb, 255:red, 74; green, 144; blue, 226 }  ,draw opacity=1 ][fill={rgb, 255:red, 74; green, 144; blue, 226 }  ,fill opacity=1 ][line width=1.5]      (0, 0) circle [x radius= 4.36, y radius= 4.36]   ;
%Straight Lines [id:da5311815373287172] 
\draw [color={rgb, 255:red, 208; green, 2; blue, 27 }  ,draw opacity=1 ][line width=1.5]    (245.76,276.16) -- (345.76,170.16) ;
\draw [shift={(345.76,170.16)}, rotate = 313.33] [color={rgb, 255:red, 208; green, 2; blue, 27 }  ,draw opacity=1 ][fill={rgb, 255:red, 208; green, 2; blue, 27 }  ,fill opacity=1 ][line width=1.5]      (0, 0) circle [x radius= 4.36, y radius= 4.36]   ;
\draw [shift={(245.76,276.16)}, rotate = 313.33] [color={rgb, 255:red, 208; green, 2; blue, 27 }  ,draw opacity=1 ][fill={rgb, 255:red, 208; green, 2; blue, 27 }  ,fill opacity=1 ][line width=1.5]      (0, 0) circle [x radius= 4.36, y radius= 4.36]   ;
%Shape: Axis 2D [id:dp5365261394213392] 
\draw  (145.46,223.16) -- (444.96,223.16)(295.76,88.05) -- (295.76,362.69) (437.96,218.16) -- (444.96,223.16) -- (437.96,228.16) (290.76,95.05) -- (295.76,88.05) -- (300.76,95.05)  ;
%Shape: Axis 2D [id:dp4432004058036936] 
\draw  (467.45,214.16) -- (770.95,214.16)(619.76,67.36) -- (619.76,365.78) (763.95,209.16) -- (770.95,214.16) -- (763.95,219.16) (614.76,74.36) -- (619.76,67.36) -- (624.76,74.36)  ;

% Text Node
\draw (535,279.48) node [anchor=north west][inner sep=0.75pt]    {$\textcolor[rgb]{0.96,0.65,0.14}{A}$};
% Text Node
\draw (691,134.48) node [anchor=north west][inner sep=0.75pt]    {$\textcolor[rgb]{0.49,0.83,0.13}{B}$};
% Text Node
\draw (635,217.48) node [anchor=north west][inner sep=0.75pt]    {$\textcolor[rgb]{0.82,0.01,0.11}{R}\textcolor[rgb]{0.82,0.01,0.11}{_{1}}$};
% Text Node
\draw (572.26,277.98) node [anchor=north west][inner sep=0.75pt]    {$z_{2}$};
% Text Node
\draw (663,128.4) node [anchor=north west][inner sep=0.75pt]    {$z_{3}$};
% Text Node
\draw (491,277.4) node [anchor=north west][inner sep=0.75pt]    {$z_{1}$};
% Text Node
\draw (725,129.4) node [anchor=north west][inner sep=0.75pt]    {$z_{4}$};
% Text Node
\draw (576,194.48) node [anchor=north west][inner sep=0.75pt]    {$\textcolor[rgb]{0.29,0.56,0.89}{R}\textcolor[rgb]{0.29,0.56,0.89}{_{2}}$};
% Text Node
\draw (211,288.48) node [anchor=north west][inner sep=0.75pt]    {$\textcolor[rgb]{0.96,0.65,0.14}{A}$};
% Text Node
\draw (367,143.48) node [anchor=north west][inner sep=0.75pt]    {$\textcolor[rgb]{0.49,0.83,0.13}{B}$};
% Text Node
\draw (297,235.48) node [anchor=north west][inner sep=0.75pt]    {$\textcolor[rgb]{0.82,0.01,0.11}{R}\textcolor[rgb]{0.82,0.01,0.11}{_{1}}$};
% Text Node
\draw (248.26,286.98) node [anchor=north west][inner sep=0.75pt]    {$z_{2}$};
% Text Node
\draw (339,137.4) node [anchor=north west][inner sep=0.75pt]    {$z_{3}$};
% Text Node
\draw (180,285.4) node [anchor=north west][inner sep=0.75pt]    {$z_{1}$};
% Text Node
\draw (401,138.4) node [anchor=north west][inner sep=0.75pt]    {$z_{4}$};
% Text Node
\draw (256,208.48) node [anchor=north west][inner sep=0.75pt]    {$\textcolor[rgb]{0.29,0.56,0.89}{R}\textcolor[rgb]{0.29,0.56,0.89}{_{2}}$};
% Text Node
\draw (277,77.4) node [anchor=north west][inner sep=0.75pt]    {$\tau $};
% Text Node
\draw (600,62.4) node [anchor=north west][inner sep=0.75pt]    {$\tau $};
% Text Node
\draw (434,229.4) node [anchor=north west][inner sep=0.75pt]    {$x$};
% Text Node
\draw (763,221.4) node [anchor=north west][inner sep=0.75pt]    {$x$};
\end{tikzpicture}
\caption{Covariant symmetric configuration. Left panel: Density matrix of the mixed state $\rho_{AB}$,  segments $A$ and $B$ are symmetric with respect to the origin. Right panel: The complimentary parts of $A\cup B$ in the system are removed by two discs with conformal invariant boundary conditions. The annulus is  a pure state  $\psi_{AB}$. \label{fig:annulus}}
\end{figure}

To include the time dependence, referring to the Euclidean path integral region in Figure \ref{fig:annulus}, we now consider $z_i=x_i +\tau_i$ for $i=1,2,3,4$ and $\tau_i\not=0$.  If we define $Z_{ij}=z_j -z_i =R_{ij}e^{i\theta_{ij}}$ and use the analytic continuation
$z_{i}=x_{i}-\tau_{i}$ and $\bar{z}_{i}=x_{i}+\tau_{i}$ to a Lorentzian
signature, the angle $\theta_{ij}$ maps to a boost parameter $\theta_{ij}=i\kappa_{ij}$. For simplicity, we denote $R_1\equiv R_{23}$ and $R_2\equiv R_{14}$. We first address the symmetric configuration by setting
\begin{equation}
z_{1}=-z_{4},\quad z_{2}=-z_{3}.
\end{equation}
Following the similar fashion as the static case,  after removing two discs representing the regions separating $A$ and $B$, we obtain an annulus indicating a pure state $\psi_{AB}$, as illustrated in Figure \ref{fig:annulus}.  The boundaries of the annulus  are imposed with conformal invariant boundary conditions, corresponding to boundary states caused by the removed regions.

\begin{figure}[h]
\begin{centering}
\includegraphics[width=0.5\textwidth]{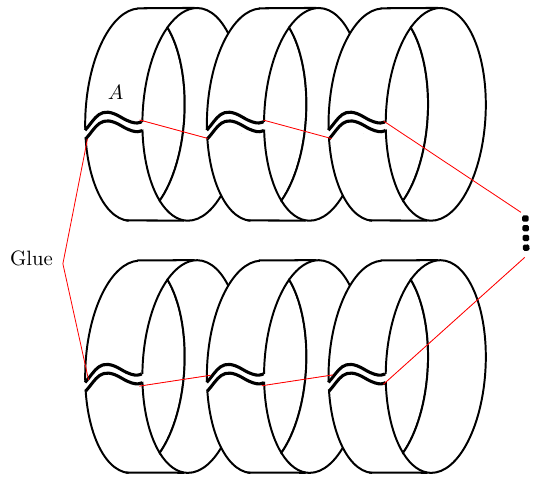}
\par\end{centering}
\caption{\label{fig:The-cut-and-glue-procedure}The cut-and-glue procedure
in the replica trick. Each annulus is cut along the subsystem $A$
and is glued with others cyclically. Red lines represent gluing operations.
Note that the resulted manifold is also an annulus.}
\end{figure}

It is helpful to define the width of the annulus
\begin{equation}
W=\log\frac{R_{2}}{R_{1}},
\end{equation}
where $R_{1}$ and $R_{2}$ denote the inner and outer radii, respectively. 
The entanglement entropy $S_{\text{vN}}(A)$ in the annulus CFT could
be defined through the R\'{e}nyi entropy:
\begin{eqnarray}
S^{(n)}(A) & = & \frac{1}{1-n}\log\mathrm{Tr}_{A}\rho_{A}^{n},\label{eq:renyi}\\
S_{\text{vN}}(A) & = & \lim_{n\rightarrow1}S^{(n)}(A).
\end{eqnarray}
$\mathrm{Tr}_{A}\rho_{A}^{n}$ is given by
\begin{equation}
\mathrm{Tr}_{A}\rho_{A}^{n}=\frac{Z_{n}}{Z_{1}^{n}}.
\end{equation}
Here, $Z_{1}$ is the partition function on the original annulus.
$Z_{n}$ is the partition function on the $n$-sheeted cover $\mathcal{M}_{n}$
obtained by cyclically gluing $n$ copies of $\mathcal{M}$ along
$A$, see Figure \ref{fig:The-cut-and-glue-procedure}. Note that
$\mathrm{Tr}_{A}\rho_{A}^{n}$ is independent of the shape of the
subsystem $A$ but depends only on the endpoints of $A$. 

By the generator of scale transformations, the annulus partition function
$Z_{1}$ can be written as \citep{Cardy:1989ir,Cardy:2004hm,Cardy:2016fqc}:
\begin{equation}
Z_{1}=e^{cW/12}\sum_{k}\langle a\vert k\rangle\langle k\vert b\rangle e^{-2\delta_{k}W}, 
\end{equation}
where $c$ is the central charge, $k$ is all allowed scalar operators
with dimensions $\delta_{k}$, and $\vert a,b\rangle$ are boundary
states on  the annulus two boundaries. The replicated manifold $\mathcal{M}_{n}$ is conformally
equivalent to an annulus with the width $W_{n}=W/n$, implying that
\begin{equation}
Z_{n}=e^{cW/12n}\sum_{k}\langle a\vert k\rangle\langle k\vert b\rangle e^{-2\delta_{k}W/n}.
\end{equation}
Since we are calculating the vacuum entanglement ($k=0$), we have
\begin{align}
Z_{1} & =e^{cW/12}\langle a\vert0\rangle\langle0\vert b\rangle,\label{eq:partition-1}\\
Z_{n} & =e^{cW/12n}\langle a\vert0\rangle\langle0\vert b\rangle,\label{eq:partition-n}
\end{align}
and 
\begin{equation}
\mathrm{Tr}_{A}\rho_{A}^{n}=e^{\frac{c}{12}(\frac{1}{n}-n)W}\left(\langle a\vert0\rangle\langle0\vert b\rangle\right)^{1-n}.\label{eq:moments}
\end{equation}
Substituting $\mathrm{Tr}_{A}\rho_{A}^{n}$ into Equation (\ref{eq:renyi}),
we obtain the R\'{e}nyi entropy
\begin{equation}
S^{(n)}(A)=\frac{c}{12}\left(1+\frac{1}{n}\right)W+g_{a}+g_{b},
\end{equation}
where $g_{a,b}=\log\langle a,b\vert0\rangle$ are the Affleck-Ludwig boundary
entropies \citep{Affleck:1991tk}.  
The boundary entropies encode  the  information from the removed subsystems.
They  are model dependent and irrelevant to the entanglement between $A$ and $B$. 
In the large $c$ limit, $g_a$ and $g_b$ are negligible compared with the first term.
So, we have
\begin{equation}
S_{\text{vN}}(A)=\lim_{n\rightarrow1}S^{(n)}(A)=\frac{c}{6}W.
\end{equation}
From the derivation, it is straightforward to see  $S_{\text{vN}}(A)=S_{\text{vN}}(B)$ since the roles of $A$ and $B$ are symmetric. Therefore, the entropy
of entanglement between $A$ and $B$ is naturally defined by the
universal term,
\begin{equation}
S_{\text{vN}}(A:B):=\frac{c}{6}W=\frac{c}{6}\log\frac{R_{2}}{R_{1}}.\label{eq:EE in width}
\end{equation}
%which agrees with Equation (\ref{eq:EE ann}).

\subsection{Generic covariant configurations}

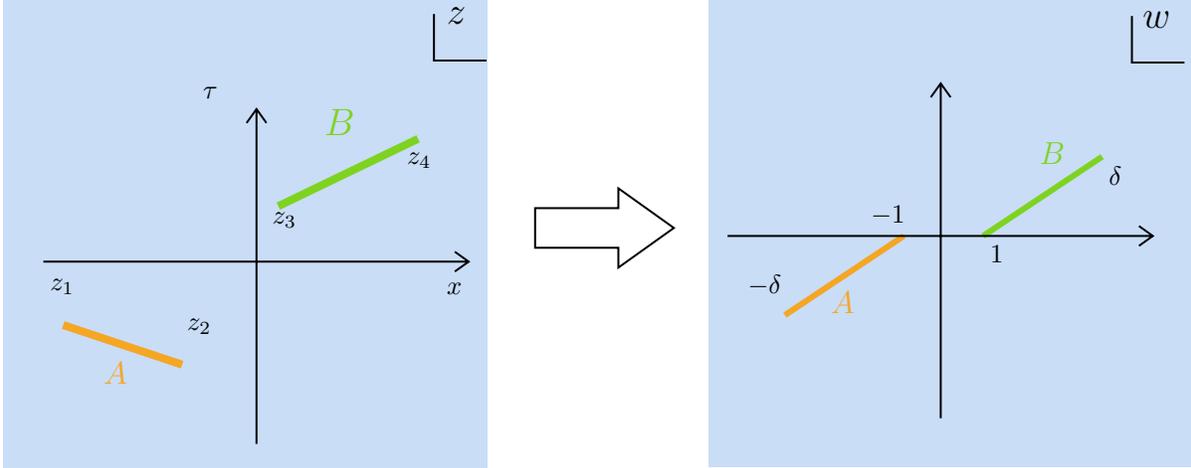
\begin{figure}[h]
\begin{centering}

\tikzset{every picture/.style={line width=0.75pt}} %set default line width to 0.75pt        

\begin{tikzpicture}[x=0.75pt,y=0.75pt,yscale=-1,xscale=1]
%uncomment if require: \path (0,675); %set diagram left start at 0, and has height of 675

%Shape: Rectangle [id:dp654839978103781] 
\draw  [color={rgb, 255:red, 0; green, 0; blue, 0 }  ,draw opacity=0 ][fill={rgb, 255:red, 74; green, 144; blue, 226 }  ,fill opacity=0.3 ] (265.49,166.38) -- (509.65,166.38) -- (509.65,403.69) -- (265.49,403.69) -- cycle ;
%Shape: Rectangle [id:dp4351846083066575] 
\draw  [color={rgb, 255:red, 0; green, 0; blue, 0 }  ,draw opacity=0 ][fill={rgb, 255:red, 74; green, 144; blue, 226 }  ,fill opacity=0.3 ] (621.49,165.38) -- (865.65,165.38) -- (865.65,402.69) -- (621.49,402.69) -- cycle ;
%Straight Lines [id:da15073286665912833] 
\draw [color={rgb, 255:red, 245; green, 166; blue, 35 }  ,draw opacity=1 ][line width=2.25]    (660,326) -- (720,286) ;
%Straight Lines [id:da9025795082297374] 
\draw [color={rgb, 255:red, 126; green, 211; blue, 33 }  ,draw opacity=1 ][line width=2.25]    (760,286) -- (820,246) ;
%Right Arrow [id:dp2550149563791666] 
\draw   (534,272) -- (576,272) -- (576,262) -- (604,282) -- (576,302) -- (576,292) -- (534,292) -- cycle ;
%Shape: Right Angle [id:dp9228921281267188] 
\draw   (509.5,197.5) -- (483,197.5) -- (483,174) ;
%Straight Lines [id:da5193518595741116] 
\draw [color={rgb, 255:red, 126; green, 211; blue, 33 }  ,draw opacity=1 ][line width=3]    (404.5,271) -- (475,237) ;
%Straight Lines [id:da1319238289495399] 
\draw [color={rgb, 255:red, 245; green, 166; blue, 35 }  ,draw opacity=1 ][line width=3]    (296,331) -- (356,351) ;
%Shape: Right Angle [id:dp22170324527568686] 
\draw   (861.5,198.5) -- (835,198.5) -- (835,175) ;
%Shape: Axis 2D [id:dp5303307475882917] 
\draw  (286,299) -- (500.5,299)(393.5,222) -- (393.5,391) (493.5,294) -- (500.5,299) -- (493.5,304) (388.5,229) -- (393.5,222) -- (398.5,229)  ;
%Shape: Axis 2D [id:dp7356320609126572] 
\draw  (631.07,286.04) -- (845.57,286.04)(738.57,209.04) -- (738.57,378.04) (838.57,281.04) -- (845.57,286.04) -- (838.57,291.04) (733.57,216.04) -- (738.57,209.04) -- (743.57,216.04)  ;

% Text Node
\draw (681.8,312.38) node [anchor=north west][inner sep=0.75pt]  [font=\large,color={rgb, 255:red, 208; green, 2; blue, 27 }  ,opacity=1 ]  {$\textcolor[rgb]{0.96,0.65,0.14}{A}$};
% Text Node
\draw (787.01,237.38) node [anchor=north west][inner sep=0.75pt]  [font=\large]  {$\textcolor[rgb]{0.49,0.83,0.13}{B}$};
% Text Node
\draw (702,269.4) node [anchor=north west][inner sep=0.75pt]    {$-1$};
% Text Node
\draw (762,289.4) node [anchor=north west][inner sep=0.75pt]    {$1$};
% Text Node
\draw (822,249.4) node [anchor=north west][inner sep=0.75pt]    {$\delta $};
% Text Node
\draw (640,304.4) node [anchor=north west][inner sep=0.75pt]    {$-\delta $};
% Text Node
\draw (488,169.4) node [anchor=north west][inner sep=0.75pt]  [font=\Large]  {$z$};
% Text Node
\draw (315,348.4) node [anchor=north west][inner sep=0.75pt]  [font=\large,color={rgb, 255:red, 245; green, 166; blue, 35 }  ,opacity=1 ]  {$A$};
% Text Node
\draw (426.22,220.37) node [anchor=north west][inner sep=0.75pt]  [font=\Large,color={rgb, 255:red, 126; green, 211; blue, 33 }  ,opacity=1 ]  {$B$};
% Text Node
\draw (288,306.4) node [anchor=north west][inner sep=0.75pt]    {$z_{1}$};
% Text Node
\draw (357,326.4) node [anchor=north west][inner sep=0.75pt]    {$z_{2}$};
% Text Node
\draw (400,272.4) node [anchor=north west][inner sep=0.75pt]    {$z_{3}$};
% Text Node
\draw (468,242.4) node [anchor=north west][inner sep=0.75pt]    {$z_{4}$};
% Text Node
\draw (839,171.4) node [anchor=north west][inner sep=0.75pt]  [font=\Large]  {$w$};
% Text Node
\draw (365,209.4) node [anchor=north west][inner sep=0.75pt]    {$\tau $};
% Text Node
\draw (488,308.4) node [anchor=north west][inner sep=0.75pt]    {$x$};

\end{tikzpicture}

\par\end{centering}
\caption{Two generic intervals determined by $z_{1},z_{2},z_{3}$ and $z_{4}$
on the $z$-plane could be conformally mapped to two symmetric intervals on the $w$-plane.\label{fig:boosted disjoint regions}}
\end{figure}
Next, we consider a generic covariant configuration where two subsystems are selected
as $A\in[z_{1},z_{2}]$ and $B\in[z_{3},z_{4}]$ without specific
constraints. Through the following conformal transformations:

\begin{equation}
v(z)=2\frac{z-z_{1}}{z_{4}-z_{1}}-1,\quad w(v)=\frac{\gamma-v}{\gamma v-1},
\end{equation}
with
\begin{equation}
\gamma=\frac{1+\alpha\beta+\sqrt{(1-\alpha^{2})(1-\beta^{2})}}{\alpha+\beta},\quad\alpha=v(z_{2}),\quad\beta=v(z_{3}),
\end{equation}
we can map a generic covariant configuration on the $z$-plane to a symmetric one
on the $w$-plane:
\begin{equation}
\left(z_{1},z_{2},z_{3},z_{4}\right)\rightarrow\left(-1,-\delta,\delta,1\right),
\end{equation}
with
\begin{equation}
\delta=\frac{1-\alpha\beta+\sqrt{(1-\alpha^{2})(1-\beta^{2})}}{\alpha-\beta},
\end{equation}
as illustrated in Figure \ref{fig:boosted disjoint regions}.
On the $w$-plane, we could apply the subtraction method, where the
inner radius is $R_{1}=1$ and the outer radius is $R_{2}=\left|\delta\right|>1$.
The corresponding entanglement entropy in this annulus is thus given
by:
\begin{eqnarray}
S_{\text{vN}}(A:B) & = & \frac{c}{6}\log\left|\delta\right|,\label{eq:Asymmetric-EE-1}\\
 & = & \frac{c}{6}\log\left|1+\frac{2(z_{1}-z_{2})(z_{3}-z_{4})}{(z_{1}-z_{4})(z_{2}-z_{3})}+2\sqrt{\frac{(z_{1}-z_{2})(z_{1}-z_{3})(z_{2}-z_{4})(z_{3}-z_{4})}{(z_{1}-z_{4})^{2}(z_{2}-z_{3})^{2}}}\right|.
 \label{eq: Covariant form}
\end{eqnarray}
In terms of the cross ratio 
\begin{equation}
\eta=\frac{z_{12}z_{34}}{z_{13}z_{24}},
\end{equation}
the covariant entanglement entropy of  mixed state $\rho_{AB}$  can be written as
\begin{eqnarray}
S_{\text{vN}}(A:B) & = & \frac{c}{6}\log\left|\frac{1+\sqrt{\eta}}{1-\sqrt{\eta}}\right|,\\
 & = & \frac{c}{12}\log\left(\frac{1+\sqrt{\eta}}{1-\sqrt{\eta}}\right)+\frac{c}{12}\log\left(\frac{1+\sqrt{\bar{\eta}}}{1-\sqrt{\bar{\eta}}}\right).
 \label{eq:Asymmetric-EE}
\end{eqnarray}
%which is correctly predicted by Equation (\ref{eq:EE in cross ratio}).
%Therefore, as discussed in Section \ref{sec:Covariant-entanglement-entropy},
%the generic bipartite system is \emph{conformally equivalent} to another symmetric bipartite system. 
%of the mixed state $\rho_{AB}$ in the covariant form.

Using Equation (\ref{eq:Asymmetric-EE}), it is straightforward
to obtain the result for a finite temperature system or a finite size system. 
To this end, one  only
needs to apply the following conformal transformations to $z$ while leaving the
form of the equation (\ref{eq:Asymmetric-EE}) invariant:
\begin{eqnarray}
\mathrm{Finite\;temperature:} &\quad  & z=\exp\frac{2\pi w}{\beta},\nonumber \\
\mathrm{Finite\;size:} & \quad & z=\tan\frac{\pi u}{L}.
\end{eqnarray}
Therefore, the covariant mixed state entanglement entropy for a finite temperature system  or a finite
size system is given by:
\begin{equation*}
    S_{\mathrm{vN}}\left(A:B\right)  = \frac{c}{12}\log\left(\frac{1+\sqrt{\eta}}{1-\sqrt{\eta}}\right)+\frac{c}{12}\log\left(\frac{1+\sqrt{\bar{\eta}}}{1-\sqrt{\bar{\eta}}}\right),
\end{equation*}
with
\begin{eqnarray}
\mathrm{Finite\;temperature:} & \quad  & \eta=\frac{\sinh\left(\pi w_{12}/\beta\right)\sinh\left(\pi w_{34}/\beta\right)}{\sinh\left(\pi w_{13}/\beta\right)\sinh\left(\pi w_{24}/\beta\right)},\nonumber \\
\mathrm{Finite\;size:} & \quad  & \eta=\frac{\sin\left(\pi u_{12}/L\right)\sin\left(\pi u_{34}/L\right)}{\sin\left(\pi u_{13}/L\right)\sin\left(\pi u_{24}/L\right)},
\end{eqnarray}
respectively.

\subsection{Two gauge parameters}

Notably, two generic intervals are uniquely fixed by eight real
numbers. However, when considering the entropy of entanglement between
two generic intervals, there are two gauge parameters. To see this,
it suffices to consider the symmetric case,
\begin{equation}
z_{1}=-z_{4},\quad z_{2}=-z_{3},
\end{equation}
where the corresponding entanglement entropy is given by (\ref{eq:EE in width}):
\begin{equation}
S_{\text{vN}}(A:B)=\frac{c}{6}\log\frac{\left|z_{1}-z_{4}\right|}{\left|z_{2}-z_{3}\right|}.
\end{equation}
Note that $S_{\text{vN}}(A:B)$ is invariant under the following \emph{local}
transformations:
\begin{equation}
(z_{1},z_{2},z_{3},z_{4})\rightarrow(z_{1}e^{\text{i}\theta},z_{2},z_{3},z_{4}e^{\text{i}\theta}),\quad(z_{1},z_{2},z_{3},z_{4})\rightarrow(z_{1},z_{2}e^{\text{i}\theta^{\prime}},z_{3}e^{\text{i}\theta^{\prime}},z_{4}),
\end{equation}
with $\theta,\theta^{\prime}$ two real numbers. Up to these local
transformations, two intervals in different configurations that cannot
be related by global conformal transformations, may still share the
same entanglement entropy. In other words, even though two generic
intervals are determined by eight real numbers, two of them are gauge
redundancy. So, there are  really only six independent parameters for generic 
covariant $S_{\text{vN}}(A:B)$.

\section{Holographic dual of the mixed state entanglement entropy\label{sec:Holography}}

For completeness, we briefly discuss the bulk dual of the entanglement entropy $S_{\text{vN}}(A:B)$, 
i.e. the covariant EWCS,  which is also addressed
in \citep{KumarBasak:2021lwm, Basu:2022nds,Wen:2022jxr}. 
The two gauge parameters discussed above play a crucial role when considering the bulk dual of the covariant entanglement entropy. 
In the AdS$_{3}$ bulk, the covariant EWCS is
determined by two endpoints, which are fixed by six real numbers. 
This precisely matches the six independent real numbers in $S_{\text{vN}}(A:B)$ after fixing
the two gauge parameters. 
Thus, a one-to-one correspondence between bulk points and boundary points
can be established.

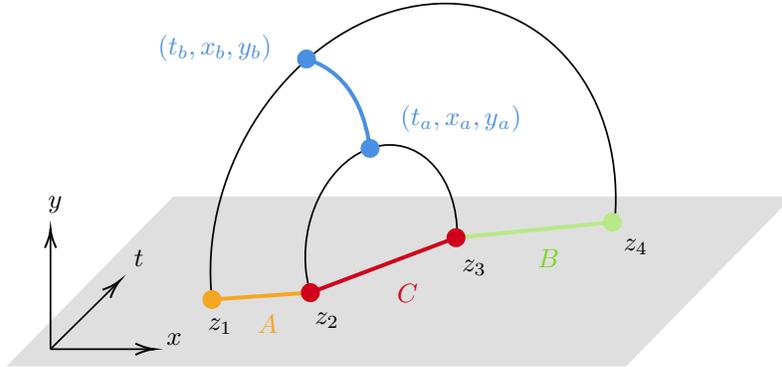
\begin{figure}[h]
	\begin{centering}

		\tikzset{every picture/.style={line width=0.65pt}} %set default line width to 0.75pt        
		
		\begin{tikzpicture}[x=0.65pt,y=0.65pt,yscale=-1,xscale=1]
			
			%Shape: Rectangle [id:dp778520689518716] 
			\draw  [color={rgb, 255:red, 255; green, 255; blue, 255 }  ,draw opacity=0 ][fill={rgb, 255:red, 155; green, 155; blue, 155 }  ,fill opacity=0.32 ] (184.5,131.08) -- (545.04,131.08) -- (447.9,230) -- (87.36,230) -- cycle ;
			%Straight Lines [id:da08293245667299809] 
			\draw    (113,220) -- (113,152.08) ;
			\draw [shift={(113,150.08)}, rotate = 90] [color={rgb, 255:red, 0; green, 0; blue, 0 }  ][line width=0.75]    (10.93,-3.29) .. controls (6.95,-1.4) and (3.31,-0.3) .. (0,0) .. controls (3.31,0.3) and (6.95,1.4) .. (10.93,3.29)   ;
			%Straight Lines [id:da3877049673374007] 
			\draw    (113,220) -- (151.94,181.49) ;
			\draw [shift={(153.36,180.08)}, rotate = 135.31] [color={rgb, 255:red, 0; green, 0; blue, 0 }  ][line width=0.75]    (10.93,-3.29) .. controls (6.95,-1.4) and (3.31,-0.3) .. (0,0) .. controls (3.31,0.3) and (6.95,1.4) .. (10.93,3.29)   ;
			%Straight Lines [id:da9488959929754828] 
			\draw    (113,220) -- (171.36,220) ;
			\draw [shift={(173.36,220)}, rotate = 180] [color={rgb, 255:red, 0; green, 0; blue, 0 }  ][line width=0.75]    (10.93,-3.29) .. controls (6.95,-1.4) and (3.31,-0.3) .. (0,0) .. controls (3.31,0.3) and (6.95,1.4) .. (10.93,3.29)   ;

			%Shape: Arc [id:dp33562461540309463] 
			\draw  [draw opacity=0] (264.24,187.08) .. controls (262.8,182.4) and (261.85,177.31) .. (261.45,171.89) .. controls (259.12,140.1) and (276.94,108.95) .. (301.24,102.31) .. controls (325.55,95.66) and (347.14,116.04) .. (349.47,147.82) .. controls (349.65,150.26) and (349.71,152.69) .. (349.65,155.11) -- (305.46,159.85) -- cycle ; \draw   (264.24,187.08) .. controls (262.8,182.4) and (261.85,177.31) .. (261.45,171.89) .. controls (259.12,140.1) and (276.94,108.95) .. (301.24,102.31) .. controls (325.55,95.66) and (347.14,116.04) .. (349.47,147.82) .. controls (349.65,150.26) and (349.71,152.69) .. (349.65,155.11) ;  
			%Shape: Arc [id:dp3233135577457269] 
			\draw  [draw opacity=0] (206.96,191.08) .. controls (206.77,189.32) and (206.6,187.54) .. (206.47,185.74) .. controls (201.17,113.4) and (249.5,40.36) .. (314.41,22.62) .. controls (379.33,4.87) and (436.25,49.13) .. (441.55,121.47) .. controls (442.13,129.37) and (442.07,137.27) .. (441.42,145.11) -- (324.01,153.6) -- cycle ; \draw   (206.96,191.08) .. controls (206.77,189.32) and (206.6,187.54) .. (206.47,185.74) .. controls (201.17,113.4) and (249.5,40.36) .. (314.41,22.62) .. controls (379.33,4.87) and (436.25,49.13) .. (441.55,121.47) .. controls (442.13,129.37) and (442.07,137.27) .. (441.42,145.11) ;  
			%Straight Lines [id:da7684258082478976] 
			\draw [color={rgb, 255:red, 245; green, 166; blue, 35 }  ,draw opacity=1 ][line width=1.5]    (206.96,191.08) -- (264.24,187.08) ;
			\draw [shift={(264.24,187.08)}, rotate = 356.01] [color={rgb, 255:red, 245; green, 166; blue, 35 }  ,draw opacity=1 ][fill={rgb, 255:red, 245; green, 166; blue, 35 }  ,fill opacity=1 ][line width=1.5]      (0, 0) circle [x radius= 4.36, y radius= 4.36]   ;
			\draw [shift={(206.96,191.08)}, rotate = 356.01] [color={rgb, 255:red, 245; green, 166; blue, 35 }  ,draw opacity=1 ][fill={rgb, 255:red, 245; green, 166; blue, 35 }  ,fill opacity=1 ][line width=1.5]      (0, 0) circle [x radius= 4.36, y radius= 4.36]   ;
			%Straight Lines [id:da5847797495724225] 
			\draw [color={rgb, 255:red, 184; green, 233; blue, 134 }  ,draw opacity=1 ][line width=1.5]    (348.96,155.08) -- (439.96,146.08) ;
			\draw [shift={(439.96,146.08)}, rotate = 354.35] [color={rgb, 255:red, 184; green, 233; blue, 134 }  ,draw opacity=1 ][fill={rgb, 255:red, 184; green, 233; blue, 134 }  ,fill opacity=1 ][line width=1.5]      (0, 0) circle [x radius= 4.36, y radius= 4.36]   ;
			%Straight Lines [id:da34828782461932906] 
			\draw [color={rgb, 255:red, 208; green, 2; blue, 27 }  ,draw opacity=1 ][line width=1.5]    (264.24,187.08) -- (348.96,155.08) ;
			\draw [shift={(348.96,155.08)}, rotate = 339.31] [color={rgb, 255:red, 208; green, 2; blue, 27 }  ,draw opacity=1 ][fill={rgb, 255:red, 208; green, 2; blue, 27 }  ,fill opacity=1 ][line width=1.5]      (0, 0) circle [x radius= 4.36, y radius= 4.36]   ;
			\draw [shift={(264.24,187.08)}, rotate = 339.31] [color={rgb, 255:red, 208; green, 2; blue, 27 }  ,draw opacity=1 ][fill={rgb, 255:red, 208; green, 2; blue, 27 }  ,fill opacity=1 ][line width=1.5]      (0, 0) circle [x radius= 4.36, y radius= 4.36]   ;
			%Curve Lines [id:da06498248574443166] 
			\draw [color={rgb, 255:red, 74; green, 144; blue, 226 }  ,draw opacity=1 ][line width=1.5]    (262,51) .. controls (290.96,60.08) and (296.96,89.08) .. (298.96,103.08) ;
			\draw [shift={(298.96,103.08)}, rotate = 81.87] [color={rgb, 255:red, 74; green, 144; blue, 226 }  ,draw opacity=1 ][fill={rgb, 255:red, 74; green, 144; blue, 226 }  ,fill opacity=1 ][line width=1.5]      (0, 0) circle [x radius= 4.36, y radius= 4.36]   ;
			\draw [shift={(262,51)}, rotate = 17.41] [color={rgb, 255:red, 74; green, 144; blue, 226 }  ,draw opacity=1 ][fill={rgb, 255:red, 74; green, 144; blue, 226 }  ,fill opacity=1 ][line width=1.5]      (0, 0) circle [x radius= 4.36, y radius= 4.36]   ;
			
			% Text Node
			\draw (110,129.4) node [anchor=north west][inner sep=0.75pt]    {$y$};
			% Text Node
			\draw (160,160.4) node [anchor=north west][inner sep=0.75pt]    {$t$};
			% Text Node
			\draw (179,209.4) node [anchor=north west][inner sep=0.75pt]    {$x$};
			% Text Node
			\draw (232,198.4) node [anchor=north west][inner sep=0.75pt]    {$\textcolor[rgb]{0.96,0.65,0.14}{A}$};
			% Text Node
			\draw (395,160.4) node [anchor=north west][inner sep=0.75pt]    {$\textcolor[rgb]{0.49,0.83,0.13}{B}$};
			% Text Node
			\draw (313,180.4) node [anchor=north west][inner sep=0.75pt]    {$\textcolor[rgb]{0.82,0.01,0.11}{C}$};
			% Text Node
			\draw (265.26,196.9) node [anchor=north west][inner sep=0.75pt]    {$z_{2}$};
			% Text Node
			\draw (351,166.32) node [anchor=north west][inner sep=0.75pt]    {$z_{3}$};
			% Text Node
			\draw (203,200.32) node [anchor=north west][inner sep=0.75pt]    {$z_{1}$};
			% Text Node
			\draw (445,154.32) node [anchor=north west][inner sep=0.75pt]    {$z_{4}$};
			% Text Node
			\draw (174,35.4) node [anchor=north west][inner sep=0.75pt]    {$\textcolor[rgb]{0.29,0.56,0.89}{( t_{b} ,x_{b} ,y_{b})}$};
			% Text Node
			\draw (315,76.4) node [anchor=north west][inner sep=0.75pt]    {$\textcolor[rgb]{0.29,0.56,0.89}{(}\textcolor[rgb]{0.29,0.56,0.89}{t}\textcolor[rgb]{0.29,0.56,0.89}{_{a}}\textcolor[rgb]{0.29,0.56,0.89}{,x}\textcolor[rgb]{0.29,0.56,0.89}{_{a}}\textcolor[rgb]{0.29,0.56,0.89}{,y}\textcolor[rgb]{0.29,0.56,0.89}{_{a}}\textcolor[rgb]{0.29,0.56,0.89}{)}$};

		\end{tikzpicture}
		\par\end{centering}
	\caption{Two generic intervals on the boundary and their corresponding EWCS
		in the AdS$_{3}$ bulk.\label{fig:bulk points}}
\end{figure}

For example, we could choose $(z_{1},z_{2},z_{3},z_{4})=(0,1,3,z_{4})$.
Following \citep{Wen:2022jxr}, the covariant EWCS is determined by
two endpoints $\left(t_{a},x_{a},y_{a}\right)$ and $\left(t_{b},x_{b},y_{b}\right)$
through the following transformations:
\begin{eqnarray}
t_{a} & = & 0,\nonumber \\
x_{a} & = & 1+\frac{2}{1+9\sqrt{\eta\bar{\eta}}},\nonumber \\
y_{a} & = & \left(\frac{36\sqrt{\eta\bar{\eta}}}{\left(9\sqrt{\eta\bar{\eta}}+1\right)^{2}}\right)^{1/2},
\end{eqnarray}
and
\begin{eqnarray}
t_{b} & = & -\frac{3\left(\eta-\bar{\eta}\right)}{9\eta\bar{\eta}-3\bar{\eta}+4\sqrt{\eta\bar{\eta}}-3\eta+1},\nonumber \\
x_{b} & = & \frac{3\left(3\eta\bar{\eta}-2\bar{\eta}-2\eta+1\right)}{9\eta\bar{\eta}-3\bar{\eta}+4\sqrt{\eta\bar{\eta}}-3\eta+1},\nonumber \\
y_{b} & = & \left(\frac{36(\eta-1)\sqrt{\eta\bar{\eta}}\left(\bar{\eta}-1\right)}{\left(-3\bar{\eta}+\eta\left(9\bar{\eta}-3\right)+4\sqrt{\eta\bar{\eta}}+1\right)^{2}}\right)^{1/2}.
\end{eqnarray}
The relations between the four boundary points of the cross ratio
$\eta$ and the two bulk points $\left(t_{a},x_{a},y_{a}\right)$
and $\left(t_{b},x_{b},y_{b}\right)$ are illustrated in Figure \ref{fig:bulk points}.
By applying these coordinate transformations, the entanglement
entropy (\ref{eq:Asymmetric-EE})  becomes:
\begin{equation}
S_{\mathrm{vN}}\left(A:B\right)=\frac{c}{6}\mathrm{arccosh}\left(\frac{-\left(t_{a}-t_{b}\right)^{2}+\left(x_{a}-x_{b}\right)^{2}+\left(y_{a}-y_{b}\right)^{2}}{2y_{a}y_{b}}+1\right),
\end{equation}
which is precisely the geodesic length between $\left(t_{a},x_{a},y_{a}\right)$
and $\left(t_{b},x_{b},y_{b}\right)$ in the AdS$_{3}$ geometry.

\begin{figure}[h]
	\begin{centering}

		\tikzset{every picture/.style={line width=0.75pt}} %set default line width to 0.75pt        
		
		\begin{tikzpicture}[x=0.75pt,y=0.75pt,yscale=-1,xscale=1]
			%uncomment if require: \path (0,741); %set diagram left start at 0, and has height of 741
			
			%Shape: Rectangle [id:dp44592760567530865] 
			\draw  [color={rgb, 255:red, 255; green, 255; blue, 255 }  ,draw opacity=0 ][fill={rgb, 255:red, 155; green, 155; blue, 155 }  ,fill opacity=0.32 ][line width=1.5]  (635.5,315.08) -- (996.04,315.08) -- (898.9,414) -- (538.36,414) -- cycle ;
			%Straight Lines [id:da20475097249063234] 
			\draw [line width=1.5]    (564,404) -- (564,337.08) ;
			\draw [shift={(564,334.08)}, rotate = 90] [color={rgb, 255:red, 0; green, 0; blue, 0 }  ][line width=1.5]    (14.21,-4.28) .. controls (9.04,-1.82) and (4.3,-0.39) .. (0,0) .. controls (4.3,0.39) and (9.04,1.82) .. (14.21,4.28)   ;
			%Straight Lines [id:da0054203404704387115] 
			\draw [line width=1.5]    (564,404) -- (602.23,366.19) ;
			\draw [shift={(604.36,364.08)}, rotate = 135.31] [color={rgb, 255:red, 0; green, 0; blue, 0 }  ][line width=1.5]    (14.21,-4.28) .. controls (9.04,-1.82) and (4.3,-0.39) .. (0,0) .. controls (4.3,0.39) and (9.04,1.82) .. (14.21,4.28)   ;
			%Straight Lines [id:da5925850433887425] 
			\draw [line width=1.5]    (564,404) -- (621.36,404) ;
			\draw [shift={(624.36,404)}, rotate = 180] [color={rgb, 255:red, 0; green, 0; blue, 0 }  ][line width=1.5]    (14.21,-4.28) .. controls (9.04,-1.82) and (4.3,-0.39) .. (0,0) .. controls (4.3,0.39) and (9.04,1.82) .. (14.21,4.28)   ;

			%Shape: Ellipse [id:dp4850867297041682] 
			\draw  [color={rgb, 255:red, 144; green, 19; blue, 254 }  ,draw opacity=1 ][fill={rgb, 255:red, 74; green, 144; blue, 226 }  ,fill opacity=0.75 ] (677.2,364.54) .. controls (677.2,342.45) and (713.02,324.54) .. (757.2,324.54) .. controls (801.38,324.54) and (837.2,342.45) .. (837.2,364.54) .. controls (837.2,386.63) and (801.38,404.54) .. (757.2,404.54) .. controls (713.02,404.54) and (677.2,386.63) .. (677.2,364.54) -- cycle ;
			%Straight Lines [id:da5130003609337849] 
			\draw [color={rgb, 255:red, 245; green, 166; blue, 35 }  ,draw opacity=1 ][line width=1.5]    (693.67,388.72) -- (717.2,364.54) ;
			%Straight Lines [id:da5426914952850287] 
			\draw [color={rgb, 255:red, 184; green, 233; blue, 134 }  ,draw opacity=1 ][line width=1.5]    (797.2,364.54) -- (819.33,339.72) ;
			%Shape: Ellipse [id:dp0026069736279581424] 
			\draw  [color={rgb, 255:red, 208; green, 2; blue, 27 }  ,draw opacity=1 ][fill={rgb, 255:red, 255; green, 255; blue, 255 }  ,fill opacity=1 ][line width=0.75]  (717.2,364.54) .. controls (717.2,353.49) and (735.11,344.54) .. (757.2,344.54) .. controls (779.29,344.54) and (797.2,353.49) .. (797.2,364.54) .. controls (797.2,375.59) and (779.29,384.54) .. (757.2,384.54) .. controls (735.11,384.54) and (717.2,375.59) .. (717.2,364.54) -- cycle ;
			%Shape: Boxed Bezier Curve [id:dp4932983476669818] 
			\draw [color={rgb, 255:red, 208; green, 2; blue, 27 }  ,draw opacity=0.4 ][fill={rgb, 255:red, 208; green, 2; blue, 27 }  ,fill opacity=0.4 ]   (717.2,364.54) .. controls (720.67,390.05) and (791.33,391.72) .. (797.2,364.54) ;
			%Shape: Boxed Bezier Curve [id:dp7765830557511341] 
			\draw [color={rgb, 255:red, 208; green, 2; blue, 27 }  ,draw opacity=1 ][fill={rgb, 255:red, 208; green, 2; blue, 27 }  ,fill opacity=0.4 ]   (717.2,364.54) .. controls (719.67,312.38) and (796.67,313.72) .. (797.2,364.54) ;
			%Curve Lines [id:da08774213278482723] 
			\draw [color={rgb, 255:red, 144; green, 19; blue, 254 }  ,draw opacity=1 ][fill={rgb, 255:red, 144; green, 19; blue, 254 }  ,fill opacity=0.2 ]   (677.2,364.54) .. controls (684.13,416.17) and (829.82,417.56) .. (837.2,364.54) ;
			%Shape: Boxed Bezier Curve [id:dp7454843701204185] 
			\draw [color={rgb, 255:red, 144; green, 19; blue, 254 }  ,draw opacity=1 ][fill={rgb, 255:red, 144; green, 19; blue, 254 }  ,fill opacity=0.2 ]   (677.2,364.54) .. controls (682.13,256.5) and (836.13,259.26) .. (837.2,364.54) ;
			%Shape: Ellipse [id:dp32779146049457986] 
			\draw  [color={rgb, 255:red, 189; green, 16; blue, 224 }  ,draw opacity=1 ][fill={rgb, 255:red, 74; green, 144; blue, 226 }  ,fill opacity=0.7 ][line width=1.5]  (337.36,349.51) .. controls (337.36,304.32) and (374.03,267.69) .. (419.27,267.69) .. controls (464.51,267.69) and (501.18,304.32) .. (501.18,349.51) .. controls (501.18,394.69) and (464.51,431.33) .. (419.27,431.33) .. controls (374.03,431.33) and (337.36,394.69) .. (337.36,349.51) -- cycle ;
			%Shape: Ellipse [id:dp7265382247950973] 
			\draw  [color={rgb, 255:red, 208; green, 2; blue, 27 }  ,draw opacity=1 ][fill={rgb, 255:red, 255; green, 255; blue, 255 }  ,fill opacity=1 ][line width=1.5]  (370.4,349.51) .. controls (370.4,322.54) and (392.28,300.69) .. (419.27,300.69) .. controls (446.26,300.69) and (468.14,322.54) .. (468.14,349.51) .. controls (468.14,376.47) and (446.26,398.33) .. (419.27,398.33) .. controls (392.28,398.33) and (370.4,376.47) .. (370.4,349.51) -- cycle ;
			%Straight Lines [id:da5872474295354185] 
			\draw [color={rgb, 255:red, 245; green, 166; blue, 35 }  ,draw opacity=1 ][line width=2.25]    (346.21,381.65) -- (370.4,349.51) ;
			%Straight Lines [id:da7635141522006166] 
			\draw [color={rgb, 255:red, 126; green, 211; blue, 33 }  ,draw opacity=1 ][line width=2.25]    (468.14,349.51) -- (493.97,316.61) ;
			%Straight Lines [id:da6489850222200914] 
			\draw    (419.27,449.51) -- (419.27,251.51) ;
			\draw [shift={(419.27,249.51)}, rotate = 90] [color={rgb, 255:red, 0; green, 0; blue, 0 }  ][line width=0.75]    (10.93,-3.29) .. controls (6.95,-1.4) and (3.31,-0.3) .. (0,0) .. controls (3.31,0.3) and (6.95,1.4) .. (10.93,3.29)   ;
			%Straight Lines [id:da5091074588756856] 
			\draw    (319.27,349.51) -- (517.27,349.51) ;
			\draw [shift={(519.27,349.51)}, rotate = 180] [color={rgb, 255:red, 0; green, 0; blue, 0 }  ][line width=0.75]    (10.93,-3.29) .. controls (6.95,-1.4) and (3.31,-0.3) .. (0,0) .. controls (3.31,0.3) and (6.95,1.4) .. (10.93,3.29)   ;
			%Straight Lines [id:da8968306262384195] 
			\draw    (419.27,349.51) -- (450.92,319.32) ;
			\draw [shift={(453.09,317.25)}, rotate = 136.36] [fill={rgb, 255:red, 0; green, 0; blue, 0 }  ][line width=0.08]  [draw opacity=0] (8.93,-4.29) -- (0,0) -- (8.93,4.29) -- cycle    ;
			%Straight Lines [id:da7617726289949881] 
			\draw    (419.27,349.51) -- (376.77,286.74) ;
			\draw [shift={(375.09,284.25)}, rotate = 55.9] [fill={rgb, 255:red, 0; green, 0; blue, 0 }  ][line width=0.08]  [draw opacity=0] (8.93,-4.29) -- (0,0) -- (8.93,4.29) -- cycle    ;
			
			% Text Node
			\draw (686,360.4) node [anchor=north west][inner sep=0.75pt]    {$\textcolor[rgb]{0.96,0.65,0.14}{A}$};
			% Text Node
			\draw (809.67,352.73) node [anchor=north west][inner sep=0.75pt]    {$\textcolor[rgb]{0.49,0.83,0.13}{B}$};
			% Text Node
			\draw (630,393.4) node [anchor=north west][inner sep=0.75pt]    {$x$};
			% Text Node
			\draw (611,344.4) node [anchor=north west][inner sep=0.75pt]    {$\tau $};
			% Text Node
			\draw (561,313.4) node [anchor=north west][inner sep=0.75pt]    {$y$};
			% Text Node
			\draw (752,350.4) node [anchor=north west][inner sep=0.75pt]    {$Q_{a}$};
			% Text Node
			\draw (749,287.4) node [anchor=north west][inner sep=0.75pt]    {$Q_{b}$};
			% Text Node
			\draw (869.73,262) node [anchor=north west][inner sep=0.75pt]   [align=left] {AdS};
			% Text Node
			\draw (339.36,352.91) node [anchor=north west][inner sep=0.75pt]    {$\textcolor[rgb]{0.96,0.65,0.14}{A}$};
			% Text Node
			\draw (483.26,331.4) node [anchor=north west][inner sep=0.75pt]    {$\textcolor[rgb]{0.49,0.83,0.13}{B}$};
			% Text Node
			\draw (511.86,353.27) node [anchor=north west][inner sep=0.75pt]    {$x$};
			% Text Node
			\draw (403.82,240.14) node [anchor=north west][inner sep=0.75pt]    {$\tau $};
			% Text Node
			\draw (441,328.38) node [anchor=north west][inner sep=0.75pt]    {$r_{a}$};
			% Text Node
			\draw (389,322.38) node [anchor=north west][inner sep=0.75pt]    {$r_{b}$};

		\end{tikzpicture}
		
		\par\end{centering}
	\caption{BCFT defined on an annulus (left) and its gravity dual (right).\label{fig:AdS/BCFT}}
\end{figure}
The subtraction approach effectively situates the BCFT on an annulus, which can be realized within the (Euclidean) AdS/BCFT correspondence \citep{Takayanagi:2011zk,Fujita:2011fp}. For an annular CFT, the gravity dual corresponds to the region bounded by two end-of-the-world branes $Q_{a,b}$ in AdS$_3$ space with the metric
\begin{equation}
	ds^2=\frac{d\tau^2+dx^2+dy^2}{y^2}
\end{equation}
as depicted in Figure \ref{fig:AdS/BCFT}. For simplicity, we assume both branes $Q_{a,b}$ are tensionless ($T_{a,b}=0$), implying vanishing boundary entropies for the annular CFT:
\begin{equation}
	g_{a,b}=\log\langle a,b\vert0\rangle=0.
\end{equation}
Consequently, the gravity dual of the BCFT on the annulus is the region in AdS$_3$ defined by
\begin{equation}
	r_a \le \sqrt{x^2+\tau^2+y^2} \le r_b,
\end{equation}
where $r_a$ and $r_b$ denote the inner and outer radii of the annulus. In this symmetric configuration, the corresponding EWCS always terminates on the branes $Q_{a,b}$. Applying conformal transformations confirms that the general covariant EWCS remains anchored to $Q_{a,b}$.

\section{Renormalization group flow induced entanglement entropy\label{sec:Renormalization-group-flow}}

In this section, we develop a more abstract but powerful approach
to study the mixed state entanglement entropy $S_{\text{vN}}$. By recognizing
that the essence of the subtraction method lies in introducing  finite
regulators, it is natural to define $S_{\text{vN}}$ by directly using
the RG flow equation in a finite region. 

In local quantum field theories (QFTs), the entanglement entropy for
an entangling surface $\Sigma$ is typically invariant under a global
rescaling of all dimensional parameters. Especially for a CFT the
universal part\footnote{Thereafter, we disregard the non-universal terms and refer to $S_{\text{EE}}$
as its universal part.} of entanglement entropy is invariant under conformal transformations,
allowing us to calculate entanglement entropy through the trace anomaly
\citep{Ryu:2006ef}. Consider a classically scale-invariant theory
living on a $d$-dimensional manifold $\mathcal{M}$ with the metric
$ds^{2}=\gamma_{ab}dx^{a}dx^{b}$, where the trace anomaly is given
by 
\begin{equation}
\left\langle T_{a}^{a}\right\rangle =2\frac{\gamma^{ab}}{\sqrt{\gamma}}\frac{\delta}{\delta\gamma^{ab}}\log Z_{\text{CFT}}.
\end{equation}
In this set-up, the Callan-Symanzik RG equation is known as \citep{Osborn:1991gm}:
\begin{equation}
\left[\ell\frac{\partial}{\partial\ell}-2\int_{\mathcal{M}}\gamma^{ab}\frac{\delta}{\delta\gamma^{ab}}\right]\log Z_{\text{CFT}}=0,
\end{equation}
with the renormalization scale $\ell$ that has the Length dimension.
Incorporating with the replica trick \citep{Calabrese:2004eu}, and
the R\'{e}nyi entropy
\begin{equation}
S_{\text{EE}}^{(n)}=\frac{1}{1-n}\log\left[\frac{Z_{\mathcal{M}_{n}}}{\left(Z_{\mathcal{M}_{1}}\right)^{n}}\right],
\end{equation}
where $\mathcal{M}_{n}$ is the replicated manifold and $\mathcal{M}_{1}$ is
the original manifold with entangling surface $\Sigma$, we can define
a RG equation for the R\'{e}nyi entropy:
\begin{equation}
\ell\frac{\partial}{\partial\ell}S_{\text{EE}}^{(n)}=\frac{\int_{\mathcal{M}_{n}}\left\langle T_{a}^{a}\right\rangle _{\mathcal{M}_{n}}-n\int_{\mathcal{M}_{1}}\left\langle T_{a}^{a}\right\rangle _{\mathcal{M}_{1}}}{1-n}.\label{eq:renyi-equation}
\end{equation}

\subsection{Two-dimensional CFT}

When $d=2$, the trace anomaly simplifies to
\begin{equation}
\left\langle T_{a}^{a}\right\rangle =\frac{c}{24\pi}\mathcal{R},\label{eq:2D-anomaly}
\end{equation}
 with $\mathcal{R}$ as the Ricci scalar and $c$ as the central charge.
For a replicated manifold with conical singularities, it was shown
\citep{Fursaev:1995ef} that 
\begin{equation}
\int_{\mathcal{M}_{n}}\mathcal{R}^{(n)}=n\int_{\mathcal{M}_{1}}\mathcal{R}+4\pi\left(1-n\right)\int_{\Sigma}1.\label{eq:replica-R}
\end{equation}
In the two-dimensional case, the entangling surface $\Sigma$ is point-like,
and $\int_{\Sigma}1=\mathcal{A}$ represents the number of points.
In what follows we assume $\mathcal{A}=1$. Plugging equations (\ref{eq:2D-anomaly})
and (\ref{eq:replica-R}) into equation (\ref{eq:renyi-equation})
and taking the limit $n\rightarrow1$, we obtain the Callan-Symanzik
equation for the entanglement entropy 
\begin{equation}
\ell\frac{\partial}{\partial\ell}S_{\text{EE}}=\frac{c}{6}.
\end{equation}
Therefore, the RG flow induced entanglement entropy, denoted as $S_{\text{RG}}$,
is 

\noindent 
\begin{equation}
S_{\text{RG}}=\frac{c}{6}\log\frac{R_{2}}{R_{1}},\label{eq:RG_EE}
\end{equation}
where $R_{1}$ and $R_{2}$ are two renormalization scales. In the
complex coordinates $(z,\bar{z})=(x+\text{i}t,x-\text{i}t)$,
the renormalization scale (Weyl scaling) $\ell$ is identified with the
radius $r=|z|=\sqrt{x^{2}+t^{2}}$ on the complex plane. The renormalization
scale region $r\in[R_{1},R_{2}]$ actually defines an annular region
on the complex plane, which precisely is the right panel of Figure \ref{fig:annulus}, 
representing the pure state obtained from a mixed state by subtraction.
Therefore, we conclude  $S_{\text{RG}}=S_{\text{vN}}$.

\subsection{Four-dimensional CFT}

It is challenging to calculate the entanglement measure for mixed
state in higher-dimensional CFTs. However, this can be easily done
in our proposal for some particular configurations. Let us consider the holographic CFT living
on a four-dimensional spacetime $ds^{2}=-dt^{2}+dx^{2}+d\vec{x}_{2}^{2}$,
where $\ensuremath{d\vec{x}_{2}^{2}}$ represents the transverse directions.
The trace anomaly for such a CFT$_{4}$ is given by
\begin{equation}
\left\langle T_{a}^{a}\right\rangle =\frac{c}{16\pi^{2}}W^{2}-\frac{a}{16\pi^{2}}E^{2},
\end{equation}
where the central charge $c$ and $a$ are equal in our case, and
the curvature square terms are expressed as
\begin{eqnarray}
W^{2} & = & R_{abcd}R^{abcd}-2R_{ab}R^{ab}+\frac{1}{3}\mathcal{R}^{2},\nonumber \\
E^{2} & = & R_{abcd}R^{abcd}-4R_{ab}R^{ab}+\mathcal{R}^{2}.
\end{eqnarray}
If we utilize the RG equation for the $n$-th R\'{e}nyi entropy (\ref{eq:renyi-equation})
and take the $n\rightarrow1$ limit, we find 
\begin{equation}
\ell\frac{\partial}{\partial\ell}S_{\text{RG}}=a\lambda_{1}\cdot\frac{\text{Area}\left(\Sigma\right)}{\ell^{2}}+a\lambda_{2},\label{eq:4d-equation}
\end{equation}
as previously found in \citep{Ryu:2006ef}. Here, both $\lambda_{1}$
and $\lambda_{2}$ are numerical constants. We consider the particular case 
that 
the entangling surface $\Sigma$ is a two-dimensional plane of
transverse directions $d\vec{x}_{2}^{2}$, with an area denoted as
$V$. In this case, the constant $\lambda_{2}$ vanishes and Eq. (\ref{eq:4d-equation})
is simplified
\begin{equation}
\ell\frac{\partial}{\partial\ell}S_{\text{RG}}=a\lambda_{1}\cdot\frac{V}{\ell^{2}}.
\end{equation}
Integrating this equation gives us:
\begin{equation}
S_{\text{RG}}=\frac{a\lambda_{1}V}{2}\left(\frac{1}{s^{2}}-\frac{1}{(2l+s)^{2}}\right),\label{eq:4D-RG-EE}
\end{equation}
where we choose the UV cutoff to be $s$ and the IR cutoff to be $2l+s$. 

Within the context of AdS$_{d+1}$/CFT$_{d}$ correspondence, the
higher-dimensional EWCS has been investigated in AdS$_{d+1}$ bulk
\citep{Jokela:2019ebz}, where the metric is given by: 
\begin{equation}
ds^{2}=\frac{R_{\text{AdS}}^{2}}{\rho^{2}}\left(-dt^{2}+d\rho^{2}+dx^{2}+d\vec{x}_{d-2}^{2}\right),
\end{equation}
Here, $\ensuremath{d\vec{x}_{d-2}^{2}}$represents the transverse
directions, and the EWCS for two parallel strips with equal widths
$l$, separated by a distance $s$, is:
\begin{eqnarray}
E_{W} & = & \frac{VR_{\text{AdS}}^{d-1}}{4G_{N}^{(d+1)}(d-2)}\frac{2^{d-2}\pi^{\frac{d-2}{2}}\Gamma\left(\frac{d}{2(d-1)}\right)^{d-2}}{\Gamma\left(\frac{1}{2(d-1)}\right)^{d-2}}\left(\frac{1}{s^{d-2}}-\frac{1}{(2l+s)^{d-2}}\right),
\end{eqnarray}
with $V$ the area of transverse directions. For $d=4$, the EWCS
is simply
\begin{equation}
E_{W}=aV\frac{\Gamma\left(\frac{2}{3}\right)^{2}}{\Gamma\left(\frac{1}{6}\right)^{2}}\left(\frac{1}{s^{2}}-\frac{1}{(2l+s)^{2}}\right),\label{eq:4D-EWCS}
\end{equation}
where the central charge $a$ is determined to be $a=\frac{R_{\text{AdS}}^{3}\pi}{2G_{N}^{(5)}}$
\citep{Myers:2010xs,Myers:2010tj}. Remarkably, the $S_{\text{RG}}$
(\ref{eq:4D-RG-EE}) perfectly matches the EWCS (\ref{eq:4D-EWCS}),
provided that $\lambda_{1}=2\,\Gamma\left(\frac{2}{3}\right)^{2}/\Gamma\left(\frac{1}{6}\right)^{2}$,
in four-dimensional case. Note that the computations of $S_{\mathrm{RG}}$ and EWCS are for two subsystems living on a static time slice  in the context of $\mathrm{AdS}_5/\mathrm{CFT}_4$ correspondence.

\section{Conclusion\label{sec:Conclusion}}

In conclusion, we have explored the subtraction
approach and the RG flow method
for calculating the entanglement
entropy $\ensuremath{S_{\text{vN}}}$ of mixed states
in CFT. By considering different configurations
and applying the subtraction method, we derived expressions for the
$S_{\text{vN}}$ of two disjoint generic intervals in CFT$_{2}$. We also established
connections between the mixed state covariant entanglement entropy and the bulk duals,
and demonstrated the significance of the gauge parameters in determining
the correspondence. Finally, we developed the RG flow approach to calculate
some universal parts in 
$S_{\text{vN}}$ and validated it within the AdS$_{5}$/CFT$_{4}$
correspondence. 
%Significantly, we observe a precise alignment between
%$S_{\text{vN}}$ in CFTs and the entanglement wedge cross section in AdS spaces.

Our proposal demonstrates that in CFT$_{2}$, the covariant mixed state entanglement
entropy can be calculated exactly. Therefore,  with the same fashion presented in \citep{Jiang:2024xcy}, the dual full spacetime geometry and dynamics, that is, the Einstein equation, can be derived from CFT$_{2}$.

\vspace*{3.0ex}
\begin{acknowledgments}
\paragraph*{Acknowledgments.} 
This work is supported in part by NSFC (Grant No. 12105191, 12275183 and 12275184).
\end{acknowledgments}

\bibliographystyle{unsrturl}
\bibliography{ref}

\end{document}